\documentclass[aps, 10pt
]{revtex4-1}

\usepackage{etex} 

\usepackage{amsmath, amssymb, arydshln,  commath, bm, empheq}
\usepackage{array}
\allowdisplaybreaks[2]
\usepackage[]{graphicx}
\DeclareGraphicsExtensions{.eps,.jpg,.png,.pdf}
\usepackage[caption=false, font=footnotesize, subrefformat=parens, labelformat=parens, belowskip=-0pt, aboveskip=-0pt,position=top]{subfig}
\usepackage{dblfloatfix}
\usepackage{booktabs}
\usepackage[toc,acronym,xindy]{glossaries} 
\newglossary[nlg]{notation}{not}{ntn}{Notation}

\newglossaryentry{sdd}{
	name={\ensuremath{\sigma_d}},
	description={standard deviation diagonal elements},}
\newglossaryentry{sdo}{
	name={\ensuremath{\sigma_o}},
	description={standard deviation off-diagonal elements},}
\newglossaryentry{cpf}{
	name={\ensuremath{\Delta}},
	description={example},} 
\newglossaryentry{a}{
	name={\ensuremath{a}}, 
	description={example},
	sort={a},}
\newglossaryentry{bsli0}{
	name={\ensuremath{\text{I}_0}}, 
	description={modified Bessel function of first kind of order 0},
	sort={b},}
\newglossaryentry{xij}{
	name={\ensuremath{\text{x}_{ij}}}, 
	description={matrix element},
	sort={x},}
\newglossaryentry{x12}{
	name={\ensuremath{\text{x}_{12}}}, 
	description={matrix element},
	sort={x},}
\newglossaryentry{yij}{
	name={\ensuremath{\text{y}_{ij}}}, 
	description={matrix element},
	sort={y},}
\newglossaryentry{y12}{
	name={\ensuremath{\text{y}_{12}}}, 
	description={matrix element},
	sort={y},}
\newglossaryentry{E}{
	name={\ensuremath{\text{E}}}, 
	description={diagonal matrix of eigenvalues},
	sort={E},}
\newglossaryentry{Ei}{
	name={\ensuremath{\text{E}_i}}, 
	description={first eigenvalue},
	sort={E},}
\newglossaryentry{Ej}{
	name={\ensuremath{\text{E}_j}}, 
	description={first eigenvalue},
	sort={E},}
\newglossaryentry{E1}{
	name={\ensuremath{\text{E}_1}}, 
	description={first eigenvalue},
	sort={E},}
\newglossaryentry{E2}{
	name={\ensuremath{\text{E}_2}}, 
	description={second eigenvalue},
	sort={E},}
\newglossaryentry{E3}{
	name={\ensuremath{\text{E}_3}}, 
	description={second eigenvalue},
	sort={E},}
\newglossaryentry{EN}{
	name={\ensuremath{\text{E}_N}}, 
	description={Nth eigenvalue},
	sort={E},}
\newglossaryentry{Es}{
	name={\ensuremath{f_\sigma}}, 
	description={constant involving function EllipticE},
	sort={E},}
\newglossaryentry{H}{
	name={\ensuremath{H}}, 
	description={matrix},
	sort={H},}
\newglossaryentry{Hij}{
	name={\ensuremath{H_{ij}}}, 
	description={matrix element},
	sort={H},}
\newglossaryentry{Hji}{
	name={\ensuremath{H_{ji}}}, 
	description={matrix element},
	sort={H},}
\newglossaryentry{Hii}{
	name={\ensuremath{H_{ii}}}, 
	description={diagonal matrix element},
	sort={H},}
\newglossaryentry{H11}{
	name={\ensuremath{H_{11}}}, 
	description={matrix element},
	sort={H},}
\newglossaryentry{H12}{
	name={\ensuremath{H_{12}}}, 
	description={matrix element},
	sort={H},}
\newglossaryentry{H21}{
	name={\ensuremath{H_{21}}}, 
	description={matrix element},
	sort={H},}
\newglossaryentry{H22}{
	name={\ensuremath{H_{22}}}, 
	description={matrix element},
	sort={H},}
\newglossaryentry{H33}{
	name={\ensuremath{H_{33}}}, 
	description={matrix element},
	sort={H},}
\newglossaryentry{HNN}{
	name={\ensuremath{H_{NN}}}, 
	description={matrix element},
	sort={H},}
\newglossaryentry{O}{
	name={\ensuremath{\text{O}}}, 
	description={Orthogonal Matrix},
	sort={O},}
\newglossaryentry{OT}{
	name={\ensuremath{\text{O}^T}}, 
	description={Transpose of Orthogonal Matrix},
	sort={O},}
\newglossaryentry{U}{
	name={\ensuremath{\text{U}}}, 
	description={Orthogonal Matrix},
	sort={U},}
\newacronym{goe}{GOE}{Gaussian Orthogonal Ensemble}
\newacronym{gue}{GUE}{Gaussian Unitary Ensemble}
\newacronym{gse}{GSE}{Gaussian Symplectic Ensemble}
\newacronym{jpdf}{JPDF}{Joint Probability Density Function}
\newacronym{ls}{LS}{Level Spacing}
\newacronym{nns}{NNS}{Nearest Neighbour Spacing}
\newacronym{pdf}{PDF}{Probability Density Function}
\newacronym{pd}{PD}{Phase Diagram}
\newacronym{rmt}{RMT}{Random Matrix Theory}
\newacronym{rnns}{RNNS}{Ratio of Nearest Neighbour Spacing}
\newacronym{rl}{RL}{Ratio of Level Spacing}
\newacronym{rms}{RMS}{Root Mean Square}
\makeglossaries
\newcommand\x{1.0}

\newcommand{\fgamma}[1]{\ensuremath{\Gamma\del{#1}}}
\newcommand{\ferf}[1]{\ensuremath{\text{Erf}\del{#1}}}
\newcommand{\ferfi}[1]{\ensuremath{\text{Erfi}\del{#1}}}
\newcommand{\ferfc}[1]{\ensuremath{\text{Erfc}\del{#1}}}

\newcommand{\pdfn}[2]{\ensuremath{\mathcal{N}\del{#1,#2}}}
\newcommand{\prob}[1]{\ensuremath{\text{P}\del{#1}}}
\newcommand{\probgoe}[1]{\ensuremath{\text{P}_{\text{GOE}}\del{#1}}}
\newcommand{\probgue}[1]{\ensuremath{\text{P}_{\text{GUE}}\del{#1}}}

\newcommand{\probcluster}[1]{\ensuremath{\text{P}_{\text{cluster}}\del{#1}}}

\newcommand{\mean}[1]{\ensuremath{\langle#1\rangle}}

\newcommand{\rotm}[2]{\ensuremath{\text{R}_{#1}\del{#2}}}

\newcommand{\matbb}[4]{\begin{bmatrix}{#1} & {#2} \\ {#3} & {#4}\end{bmatrix}}

\begin{document}


\title{Eigenvalue Statistics for Generalized Symmetric and Hermitian Matrices}
\author{Adway Kumar Das}\email{akd14ms147@iiserkol.ac.in}
\author{Anandamohan Ghosh}\email{anandamohan@iiserkol.ac.in}
\affiliation{
	Indian Institute of Science Education and Research Kolkata, Mohanpur, 741246 India
}

\date{\today}

\begin{abstract}
The \acrfull{nns} distributions can be computed for generalized symmetric $2 \times 2$ matrices having different variances in the diagonal and in the offdiagonal elements. Tuning the relative value of the variances we show that the distributions of the level spacings exhibit a  crossover from clustering to repulsion as in GOE. The analysis is extended to $3 \times 3$ matrices where distributions of \acrshort{nns} as well as \acrfull{rnns} show similar crossovers. We show that it is possible to calculate \acrshort{nns} distributions for Hermitian matrices ($N=2,3$) where also crossovers take place between clustering and repulsion as in GUE. For large symmetric and Hermitian matrices we use interpolation between clustered and repulsive regimes and identify phase diagrams with respect to the variances.
\end{abstract}

\pacs{Valid PACS appear here}
\maketitle

\section[Introduction]{Introduction}{\label{sect_int}}

\acrfull{rmt} has emerged as an important statistical tool to distinguish irregular and chaotic dynamics from regular and integrable dynamics of quantum systems \cite{haake2013quantum}. 
It has found applications in a variety of disciplines ranging from energy level fluctuations in nuclear physics \cite{Guhr1} to chiral phase transitions in quantum chromodynamics \cite{Verbaarschot1}
to more recent studies on many body localization and thermalization in condensed matter physics \cite{borgonovi2016quantum,d2016quantum}.
\acrfull{rmt} can predict some universal characteristics, without explicit knowledge of the Hamiltonian (or any suitable operator), solely dictated by the underlying symmetries of the dynamical system \cite{Bohigas1}. 
In this regard, the most investigated quantity is \acrfull{nns} of energy levels \cite{Wigner1}.
If a quantum system follows regular dynamics i.e. the system is in integrable domain, the \acrfull{pdf} of \acrshort{nns} follows a Poisson distribution \cite{Berry3}, which implies level clustering. 
On the other hand, level repulsion is seen in systems having time-reversal and rotational invariance (or time-reversal invariant system with integer spin and broken rotational symmetry) and is described by Wigner surmise for \acrfull{goe} with Dyson index $\beta=1$ \cite{Bohigas1}.
The corresponding classical counterparts of these systems show chaos, hence level repulsion can be considered to be a signature of quantum chaos \cite{haake2013quantum}.
Two more ensembles are introduced following group theoretical arguments \cite{Dyson1}, namely \acrfull{gue} (broken time-reversal symmetry) for $\beta=2$ and \acrfull{gse} (time-reversal invariant system with half-integer spin and broken rotational symmetry) for $\beta=4$. 
Universal features in \acrshort{nns} distributions have been observed on suitable normalization and {\it unfolding} of the spectrum. 
Recently a simpler method, avoiding numerical issues of the unfolding procedure, has been proposed to compute distributions of \acrfull{rnns} which also show universal features of random matrix ensembles \cite{Atas1}.

All of the above ensembles are pure in the sense that they provide description of dynamical systems that are either regular or chaotic. 
However, level statistics found in many physical systems often indicate intermediate states that are neither purely integrable or chaotic \cite{bogomolny1999models,izrailev1989intermediate} and such intermediate statistics in level fluctuations has also been experimentally observed  \cite{Rosenzweig1, Alt1, Abdul-Magd1, Robnik1}. 
In order to describe the mixed dynamics, many phenomenological models have been suggested for e.g., Brody distribution \cite{Brody1}, Berry-Robnik distribution \cite{Berry2}, \acrshort{goe}-\acrshort{gue} transition \cite{Pandey1}, cross-over between Poisson-\acrshort{goe}-\acrshort{gue} \cite{Schierenberg1, Schweiner1,vivo2008invariant}, etc. 
These models explain the transition/crossover based on \acrshort{nns} of eigenvalues with a few recent studies based on \acrshort{rnns} distributions \cite{Chavda1}. 
Typical approaches in these set-up are to consider additive random matrix models where a particular symmetry is broken perturbatively by tuning an interpolating parameter.
There are also some limitations in these phenomenological models, for example, the derivative of the Brody distribution diverges at zero energy \cite{Grammaticos1}, the Berry-Robnik distribution does not give level repulsion in chaotic regime \cite{Huu-Tai1}, absence of scaling property \cite{Cheon1}. 

The intermediate statistics existing between limiting ensembles can be accessed by tuning a transition parameter but it lacks any physical interpretation. 
A relevant way to explore the mixed features is to consider generalized random matrices and investigate the possibility of any transition by tuning the statistical properties of the random matrix elements.
Generalization have been possible for symmetric Gaussian matrices where
diagonal and off-diagonal elements are drawn from normal distributions with different mean and variances  \cite{Huu-Tai1, Berry1}.
In this paper, we have studied the crossover between level repulsion and clustering by tuning the relative variance of the diagonal ($\gls{sdd}$) and the off-diagonal ($\gls{sdo}$)  elements of symmetric random matrices. 
We extend the analysis for $3\times{3}$ matrices by proposing an ansatz for the eigenvalue distribution and obtain analytical expressions for the NNS distributions as well as the RNNS distributions
that agrees with simulation results.
We obtain interpolating functions, parameterized by tunable parameters which are numerically estimated, and different phases are identified in $\gls{sdd}$-$\gls{sdo}$ plane.
We show that the analysis is also applicable to generalized Hermitian matrices and exact results are obtained for $N=2$ again showing crossover with respect to the variances. 
For higher values of $N$ we have relied on numerical data and demonstrate that the crossover from level clustering to repulsion is a generic feature of generalized random matrices.

\section[Symmetric matrices]{Symmetric matrices}{\label{sect_sym}}

Let us consider a matrix $H$ composed of $m$-independent entries $x_1,x_2,\dots,x_m$ each drawn from a probability distribution $P(x)$ implying that, 
\begin{align}
	\label{eq_fullpdf}
	\prob{\gls{H}}=\prod_{i=1}^{m}\prob{x_i}.
\end{align}
We are interested in diagonalizing  matrix $H = \Theta^{-1} \mathcal{E} \Theta$
and obtaining the joint probability distribution (JPDF) of eigenvalues, 
$\prob{\mathcal{E}}=\prob{\gls{E1},\gls{E2},\dots,\gls{EN}}$.
If we consider that the matrix elements of eigenfunctions $\Theta$ are parameterized as 
$\cbr{\theta_1,\theta_2,\dots,\theta_M}$ then the transformation from matrix space to eigenspace necessitates
\begin{align}
	\label{eq_mattoeig}
	\prob{x_1,\dots,x_m}\prod_{i}^{m}dx_i &= f\del{\gls{E1},\dots,\gls{EN},\theta_1,\dots,\theta_M}|J|\prod_{j}^{N}dE_j\prod_{k}^{M}d\theta_k
\end{align}
where, $J$ is the Jacobian of the transformation. To find the JPDF of eigenvalues we need to integrate over $\theta_i$
\begin{align}
	\label{eq_pdfhtoe}
	\prob{\mathcal{E}} &= |J|\int{d\theta_1}\dots\int{d\theta_M}f\del{\gls{E1},\dots,\gls{EN},\theta_1,\dots,\theta_M}
\end{align}
which is suitably normalized such that $\int{d\mathcal{E}}~\prob{\mathcal{E}}=1$. 
If $\prob{\mathcal{E}}$ is symmetric function of its arguments, i.e. $\prob{\gls{E1},\gls{E2},\dots,\gls{EN}}=\prob{E_{i1},E_{i2},\dots,E_{iN}}$, where $\cbr{i1,i2,\dots,iN}$ are arbitrary permutations of $\cbr{1,2,\dots,N}$, then marginal \acrshort{pdf} of eigenvalue, $\prob{E}$ is given by,
\begin{align}
	\label{eq_pdfEtoe}
	\prob{E} &= \int{d\gls{E2}}\dots\int{d\gls{EN}}\prob{\mathcal{E}}.
\end{align}
In this section we consider \gls{H} as a $N{\times}N$ real symmetric matrix, $\gls{Hij}=\gls{Hji}$, 
with the diagonal and the offdiagonal entries drawn from normal distributions such that
$\gls{Hii}{\sim}\pdfn{0}{\gls{sdd}^2}$ and 
$\gls{Hij}{\sim}\pdfn{0}{\gls{sdo}^2}{\quad\forall\quad}i{\neq}j$ respectively. 
Then the density function of $H$ given in Eq.(\ref{eq_fullpdf}) can be written as
\begin{align}
	\label{eq_pdfh}
	\begin{split}
	\prob{\gls{H}} &= C\exp\del{-\sum_{i=1}^{N}\frac{\gls{H}_{ii}^2}{2\gls{sdd}^2}-\sum_{i<j}^{N}\frac{\gls{Hij}^2}{2\gls{sdo}^2}}\\
	&= C\exp\del{\del{\frac{1}{4\gls{sdo}^2}-\frac{1}{2\gls{sdd}^2}}\sum_{i}^{N}\gls{H}_{ii}^2-\frac{\text{Tr}(\gls{H}^2)}{4\gls{sdo}^2}}\qquad\sbr{\text{ where, }C = \frac{(2\pi)^{-\frac{N(N+1)}{4}}}{\gls{sdd}^{N}\gls{sdo}^{\frac{N(N-1)}{2}}}}
	\end{split}
\end{align}
For symmetric matrices, eigenvectors are orthogonal to each other and $\Theta$ becomes an orthogonal matrix, \gls{O}. 
Then, we can do a similarity transformation, $\gls{H} = \gls{OT}\mathcal{E}\gls{O}$, which implies the canonical invariance $\text{Tr}(\gls{H}^2) = \sum_{i}^{N}\gls{Ei}^2$. 
Moreover, the Jacobian $J$ is given by the Vandermonde determinant, i.e. $J\del{\gls{H}\to \cbr{\mathcal{E},\gls{O}}}=\prod_{j<k}^{N}(\gls{Ei}-\gls{Ej})$ \cite{Livan1}. 
Using the above properties, the JPDF of eigenvalues assume the form
\begin{align}
	\label{eq_mainpdfe}
	&\prob{\mathcal{E}}\propto \exp\del{-\frac{1}{4\sigma^2}\sum_{i}^{N}\gls{Ei}^2}\prod_{i<j}^{N}|\gls{Ei}-\gls{Ej}|
\end{align}
Symmetric matrices satisfying $\gls{sdo}=\sqrt{2}\gls{sdd}$ belong to Gaussian Orthogonal Ensemble (GOE) corresponding to $\beta=1$ in Wigner's surmise. For any arbitrary choices of  $\{\gls{sdo}, \gls{sdd}\}$ obtaining an analytical expression of $\prob{\mathcal{E}}$ 
is difficult and becomes harder as $N$ increases.
We now illustrate the generalization of Wigner surmise proposed for $2\times2$ matrices \cite{Huu-Tai1} and 
show that by tuning $\{\gls{sdo}, \gls{sdd}\}$ crossovers are possible 
between level repulsion and clustering of eigenvalue spectra.

\subsection{NNS distributions for $N=2$}
For $N=2$, $\Theta$ can be taken as the $2\times 2$ rotation matrix, i.e. $\gls{O}=\begin{pmatrix}
\cos\theta&\sin\theta\\-\sin\theta&\cos\theta
\end{pmatrix}$ and Eq.~\eqref{eq_mattoeig} becomes
\begin{align}
	\label{eq_trans2}
	& \matbb{\gls{H11}}{\gls{H12}}{\gls{H21}}{\gls{H22}} = \matbb{\gls{E1}\cos^2\theta+\gls{E2}\sin^2\theta}
	{(\gls{E1}-\gls{E2})\sin\theta\cos\theta}
	{(\gls{E1}-\gls{E2})\sin\theta\cos\theta}
	{\gls{E1}\sin^2\theta+\gls{E2}\cos^2\theta}.
\end{align}

\renewcommand{\x}{0.33}
\captionsetup[subfigure]{position=top, labelfont=bf,textfont=normalfont,singlelinecheck=off,justification=raggedright}
\begin{figure}[!t]
	\vspace{-10pt}
	\centering
	\subfloat[][]{\includegraphics[width=\x\textwidth, trim=210 10 230 20, clip=true]{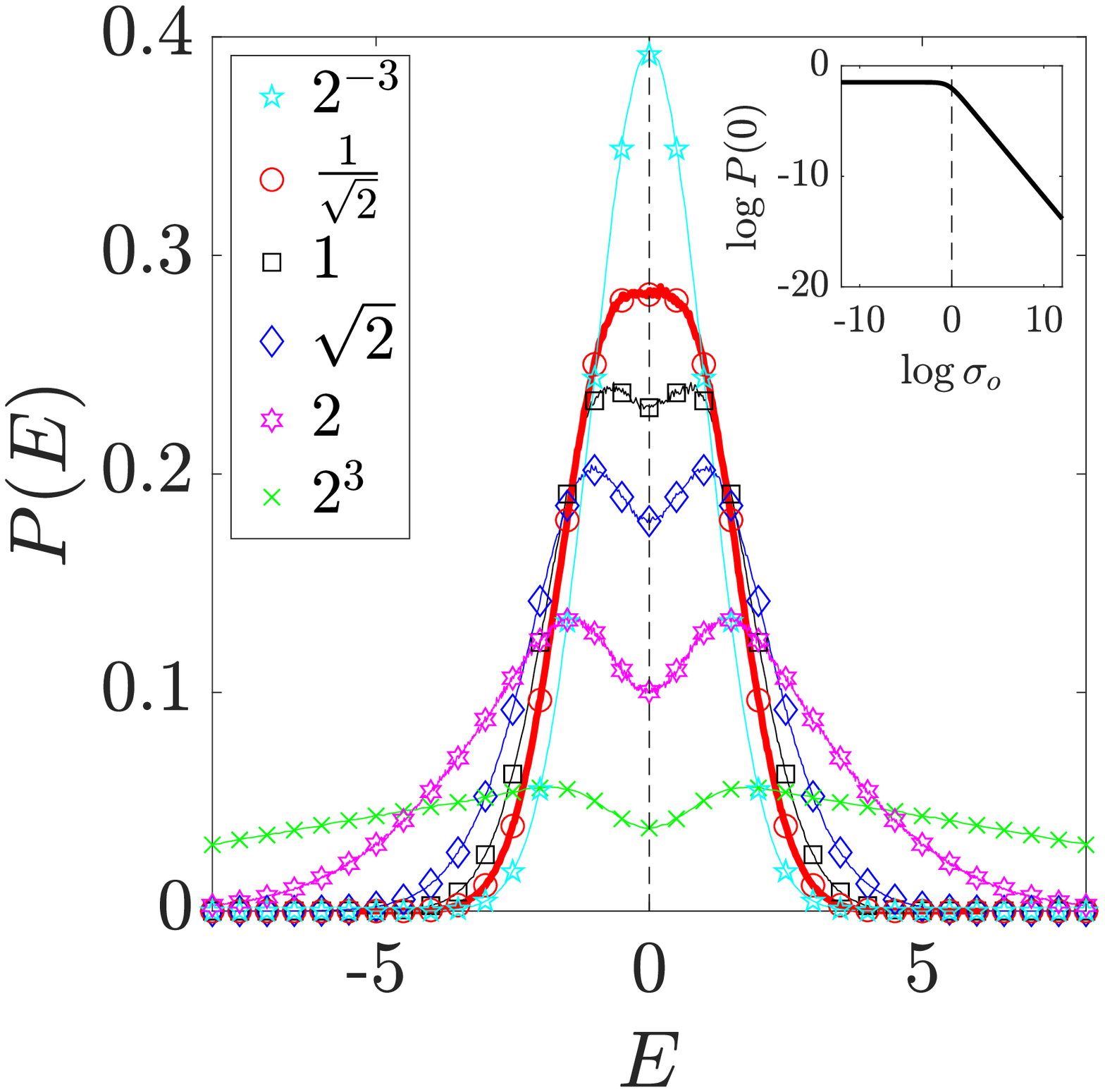}\label{eig_all}}
	\subfloat[][]{\includegraphics[width=\x\textwidth, trim=90 0 120 20, clip=true]{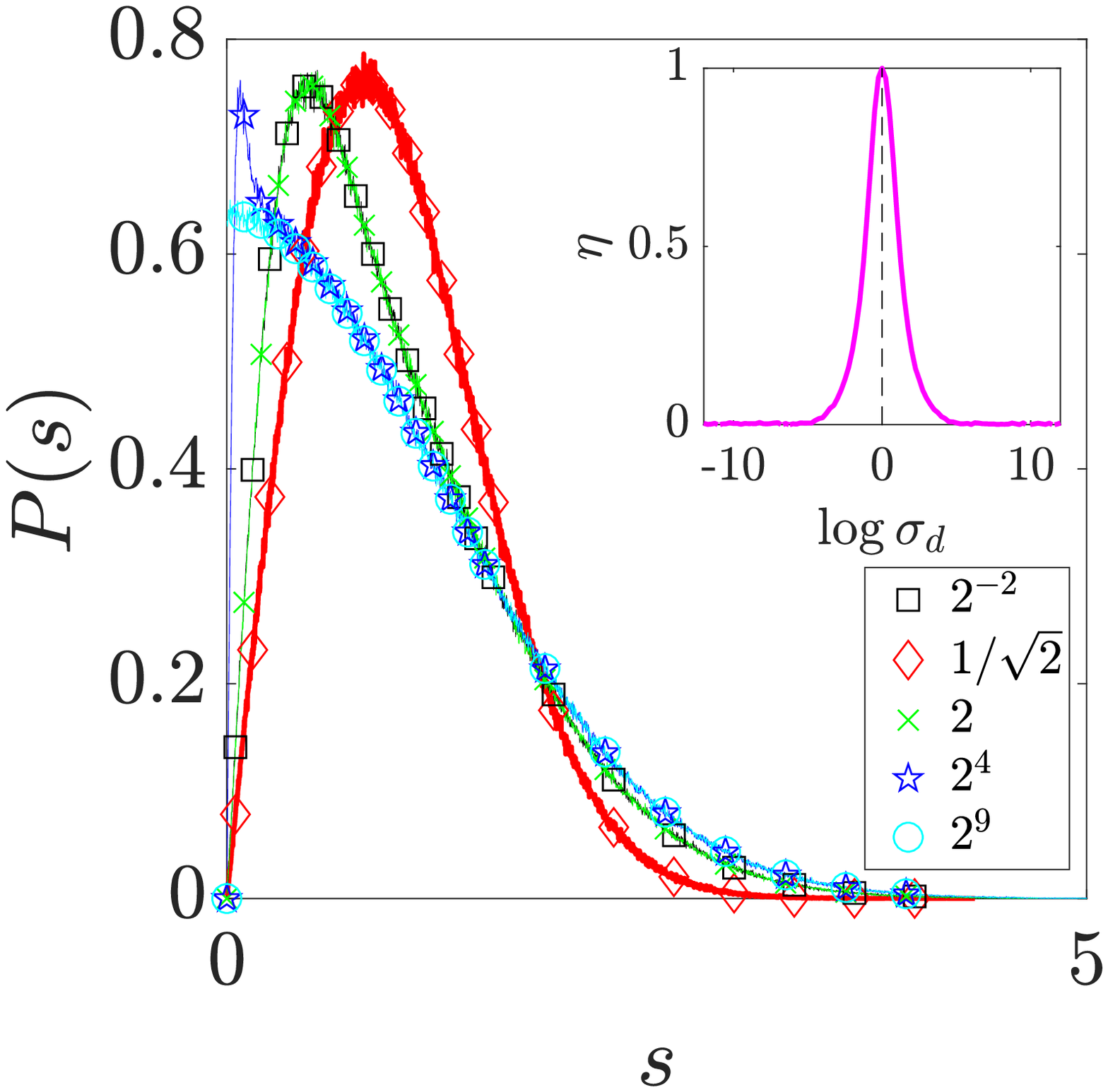}\label{ls_all}}
	\subfloat[][]{\includegraphics[width=\x\textwidth, trim=90 0 120 20, clip=true]{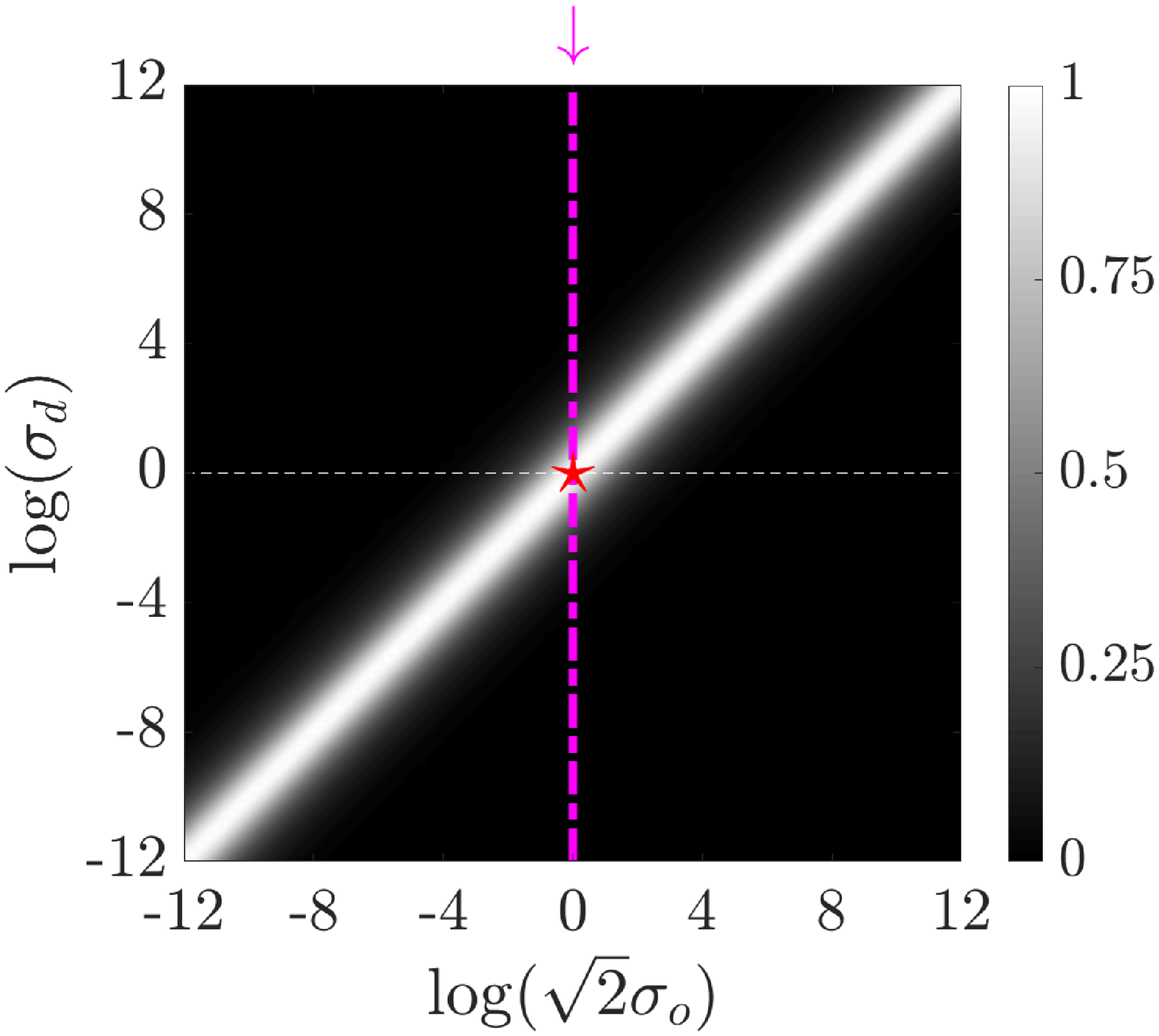}\label{pdiag_sym_2}}
	\caption[]{{\bf Results for $2\times2$ symmetric matrix}:
	\subref{eig_all} PDF of eigenvalue for different $\gls{sdo}$ with $\gls{sdd}=1$. 
	Markers denote expression in Eq.~\eqref{eq_marginalpdf} and continuous lines are simulations.
	The red line corresponds to GOE.
	Inset shows $\prob{E=0}$ vs \gls{sdo} in log-log scale. 
	\subref{ls_all} \acrshort{pdf} of  ~{\acrshort{nns}}. 
		Inset shows transition parameter $\eta$ vs $\log(\gls{sdd})$ when $\gls{sdo}=1/\sqrt{2}$. 
	\subref{pdiag_sym_2}~Phase diagram of transition from level clustering (dark) to level repulsion (light).  
	The red star indicates GOE obtained from Eq.~\eqref{eq_pdfs2}. 
	The inset in \subref{ls_all} corresponds to the dot-dashed line marked with arrow.
	In simulations averages are computed over 500000 samples.
}
	\label{sym_ls_all}
\end{figure}

Substituting $H_{ij}$ in {Eq.~\eqref{eq_pdfh}} and integrating over $\theta$ and normalization yields
quadratic Rayleigh-Rice distribution \cite{Huu-Tai1}
\begin{align}
	\label{eq_pdfe1e2}
	\begin{split}
	& \prob{\mathcal{E}} = \frac{|\gls{E1}-\gls{E2}|}{4\sqrt{2\pi}\gls{sdd}^2\gls{sdo}}\exp\del{f_{\mathcal{E}}-\sum_{i=1}^{2}\frac{\gls{Ei}^2}{2\gls{sdd}^2}}\gls{bsli0}\del{f_{\mathcal{E}}}\\
	&\text{ where, }f_{\mathcal{E}}=\sbr{\frac{1}{8\gls{sdd}^2}-\frac{1}{16\gls{sdo}^2}}(\gls{E1}-\gls{E2})^2.
	\end{split}
\end{align}
$\gls{bsli0}$ is the modified Bessel function of 1$^{st}$ kind of order 0. 
Using the above JPDF in Eq.~\eqref{eq_pdfEtoe}, we get the marginal \acrshort{pdf} of eigenvalues,
\begin{align}
	\label{eq_marginalpdf}
	\prob{E} =& \frac{1}{4\sqrt{2\pi}\gls{sdd}^2\gls{sdo}}\int_{-\infty}^{\infty}dx|x|\exp\del{-\frac{(y+2E)^2}{4\gls{sdd}^2}-\frac{y^2}{8}\del{\frac{1}{\gls{sdd}^2}+\frac{1}{2\gls{sdo}^2}}}\gls{bsli0}\del{\frac{y^2}{8}\del{\frac{1}{\gls{sdd}^2}-\frac{1}{2\gls{sdo}^2}}}.
\end{align}
The above expression can be numerically evaluated and are shown for different $\cbr{\gls{sdd},\gls{sdo}}$ in Fig.~\ref{eig_all}. 
For $\gls{sdd}=1$ and $\gls{sdo}={1}/{\sqrt{2}}$, the expression in {Eq.~\eqref{eq_pdfe1e2}} gives \acrshort{pdf} of eigenvalues for \acrshort{goe}
\begin{align}
	\label{eq_pdfe1e2goe}
	& \probgoe{\gls{E1},\gls{E2}} = \frac{|\gls{E1}-\gls{E2}|}{4\sqrt{\pi}}\exp\del{-\frac{\gls{E1}^2+\gls{E2}^2}{2}}.
\end{align}
Knowing JPDF the integration in Eq.~\eqref{eq_pdfEtoe} can be computed exactly for \acrshort{goe}  \cite{Mehta1} and we get
\begin{align}
	\label{eq_pdfegoe}
	&\probgoe{E} = \frac{1}{2\sqrt{\pi}}e^{-E^2}+\frac{1}{2\sqrt{2}}Ee^{-E^2/2}\ferf{\frac{E}{\sqrt{2}}}
\end{align}
and is shown by the red line in Fig.~\ref{eig_all}. 
Next we calculate \acrshort{pdf} of the \acrshort{nns}, $s$, as
\begin{align}
	\label{eq_pdfsunnormed}
	&\prob{s} = \int_{-\infty}^{\infty}d\gls{E1}d\gls{E2}P(\gls{E1},\gls{E2})\delta(s-|\gls{E1}-\gls{E2}|).
\end{align}
Note that to evaluate {Eq.~\eqref{eq_pdfsunnormed}}, considering the case $\gls{E1}\geq \gls{E2}$ will suffice. Due to the symmetry of Eq.~\eqref{eq_pdfe1e2}, the other possibility will yield same results. 
$\prob{s}$ is normalized and unfolded by changing the unit of \acrshort{nns} such that $\mean{s}=1$ to produce an universal expression \cite{Brody2} by demanding
\begin{align}
	\label{eq_pdfs}
	&\int_{0}^{\infty}\prob{s}~ds=1,\quad \mean{s}=\int_{0}^{\infty}\prob{s}~s~ds=1.
\end{align}

Let us define $\tilde{\sigma}=\frac{\gls{sdd}}{\sqrt{2}\gls{sdo}}$ and using {Eq.~\eqref{eq_pdfe1e2}} in {Eq.~\eqref{eq_pdfsunnormed}} and satisfying Eq.~\eqref{eq_pdfs} we get 
\begin{align}
	\label{eq_pdfs2}
	&\prob{s} = \frac{2\gls{Es}^2}{\pi\sigma} s\exp\del{-\frac{(1+\sigma^2)\gls{Es}^2s^2}{2\pi\sigma^2}}\gls{bsli0}\del{\frac{(1-\sigma^2)\gls{Es}^2s^2}{2\pi\sigma^2}}\quad\sbr{\gls{Es} = \text{EllipticE}(1-\sigma^2), \sigma = \min\cbr{\tilde{\sigma},\frac{1}{\tilde{\sigma}}}}
\end{align}
where EllipticE is complete elliptic integral \cite{Berry1} .
If {Eq.~\eqref{eq_pdfs2}} is observed carefully, it shows the following symmetry,
\begin{align}
	\label{eq_sym2symmetry}
	\prob{s}_{\tilde{\sigma}=x} &= \prob{s}_{\tilde{\sigma}=\frac{1}{x}}\quad\forall\quad x>0
\end{align}
which implies \acrshort{nns} distributions are identical for $\gls{sdd}\gg\gls{sdo}$ as well as $\gls{sdd}\ll\gls{sdo}$.
It is important to note that $\prob{\mathcal{E}}$ in Eq.~\eqref{eq_pdfe1e2} or $\prob{E}$ in Eq.~\eqref{eq_marginalpdf} does not have the above symmetry.
For \acrshort{goe} i.e. $\gls{sdd}=\sqrt{2}\gls{sdo}$ it is easy to see that {Eq.~\eqref{eq_pdfs2}} 
reduces to the familiar
\begin{align}
	\label{eq_pdfs2goe}
	&\probgoe{s} = \frac{\pi}{2}s\exp\del{-\frac{\pi}{4}s^2}.
\end{align}
Thus level repulsion is evident for $\tilde\sigma\to 1$.
On the other hand, level clustering can be shown for $\gls{sdo}\to{0}$, where the offdiagonal elements $\gls{Hij}\approx 0\quad{\forall}{\quad}i{\neq}j$, thus $\prob{H} \approx \frac{1}{2\pi\gls{sdd}^2}\exp\del{-\sum_{i=1}^{2}\frac{\gls{Hii}^2}{2\gls{sdd}^2}}$. In this limit, the diagonal elements are the eigenvalues themselves, i.e. $\prob{\mathcal{E}}=\frac{1}{2\pi\gls{sdd}^2}\exp\del{-\sum_{i=1}^{2}\frac{\gls{Ei}^2}{2\gls{sdd}^2}}$, giving
\begin{align}
	\label{eq_pdfpoissons}
	\probcluster{s} &= \frac{2}{\pi}\exp\del{-\frac{s^2}{\pi}}
\end{align}
corresponding to clustering as $s\rightarrow 0$. However, for any small value of $\gls{sdo}$ the enumeration of $\prob{s}$ in Eq.~\eqref{eq_pdfs2} shows a sharp jump to maxima near $s=0$ and then rapidly decays to 0. 
This jump is sharper with decreasing $\gls{sdo}$ such that for all practical
purposes we consider this as level clustering described by Eq.~\eqref{eq_pdfpoissons}.
Due to the symmetry in {Eq.~\eqref{eq_pdfs2}} the observations for $\gls{sdo}\to{0}$ i.e. $\tilde{\sigma}\to{\infty}$ are also applicable for the condition $\tilde{\sigma}\to{0}$. 
 This indicates a crossover between level repulsion ($\tilde\sigma\to 1$) and level clustering as relative strengths of the variances of the diagonal and offdiagonal elements are varied.
We would also like to mention that in integrable systems the level clustering is associated with Poisson distribution, $\prob{s}=e^{-s}$, which is obtained after a local transformation, $\hat{s}=sNP_X(x)$ \cite{Livan1}, and
is realizable for large $N$ while here we are considering matrices with size $N=2$.

In order to characterize the crossover we need an empirical function for intermediate values of $\cbr{\gls{sdd},\gls{sdo}}$ similar in the spirit of Brody distribution \cite{Brody1}.
In our case, the limiting cases for level repulsion, $\prob{s}\sim{s}\exp\del{-s^2}$, and for level clustering, $\prob{s}\sim\exp\del{-s^2}$, suggest a transition function of the form $\prob{s}\sim{s}^\eta\exp\del{-s^2}$. Normalizing and unfolding this equation we get
\begin{align}
	\label{eq_phase_2s}
	&\prob{\eta; s} = \frac{2\fgamma{1+\frac{\eta}{2}}^{1+\eta}}{\fgamma{\frac{1+\eta}{2}}^{2+\eta}}s^\eta\exp\del{-\frac{4\fgamma{1+\frac{\eta}{2}}^2}{\fgamma{\frac{1+\eta}{2}}^2}s^2}.
\end{align}
The data obtained from enumeration of the expression in Eq.~\eqref{eq_pdfs2} are 
fitted to the above function and $\eta$ is estimated using trust-region algorithm \cite{Yuan1}. A value $\eta\approx1$ will imply level repulsion while $\eta\approx0$ will imply level clustering. Based on the values of $\eta$ we can construct the phase diagram as shown in Fig.~\ref{sym_ls_all}\subref{pdiag_sym_2} where dark (white) region shows level clustering (repulsion). The GOE is represented at the centre of these \acrshort{pd}. The phase diagram indicates that from the region where $\gls{sdd}=\sqrt{2}\gls{sdo}$ ($\eta\to1$) changing the difference between $\gls{sdd}$ and $\gls{sdo}$ by one order of magnitude results in a crossover ($\eta\to0$).

The choice of the distribution for elements of a symmetric matrix can be further generalized, where $\gls{H11}\sim{N(\mu_1,\sigma_1)}, \gls{H12}=\gls{H21}\sim{N(\mu_2,\sigma_2)}, \gls{H22}\sim{N(\mu_3,\sigma_3)}$ \cite{Berry1}. However, for $\mu_1=\mu_2=\mu_3=0$ we can redefine $\tilde{\sigma}=\sqrt{\sigma_1^2+\sigma_3^2}/2\sigma_2$, then again $\prob{s}$ follow {Eq.~\eqref{eq_pdfs2}}.

\subsection{ NNS distributions for $N = 3$}

In the case of $3{\times}3$ matrix, \gls{O} can be taken as generalized $3{\times}3$ rotation matrix, i.e. $\gls{O}=\rotm{x}{\theta}\rotm{y}{\phi}\rotm{z}{\psi}$, and the transformation rules can be obtained for $H_{ij}$.
However, subsequent calculations are tedious and obtaining simple expressions become impossible. 
Observing {Eq.~\eqref{eq_mainpdfe}} and {Eq.~\eqref{eq_pdfe1e2}} for $N=2$ 
we propose the following ansatz for generalized $N\times N$ symmetric matrices
\begin{align}
	\label{eq_pdfen}
	\begin{split}
	\prob{\mathcal{E}} &= C_0\exp\del{-\sum_{i=1}^{N}\frac{\gls{Ei}^2}{2\sigma_d^2}+\sum_{i<j}^{N}f_{\gls{Ei},\gls{Ej}}}\prod_{i<j}^{N}\del{|\gls{Ei}-\gls{Ej}|\gls{bsli0}\del{f_{\gls{Ei},\gls{Ej}}}},\quad f_{\gls{Ei},\gls{Ej}} =  -\frac{1}{C}\Bigg|\frac{1}{\sigma_d^2}-\frac{1}{2\sigma_o^2}\Bigg|\del{\gls{Ei}-\gls{Ej}}^2
	\end{split}
\end{align}
where the unknown constant $C$ and the normalization constant $C_0$ need to be numerically determined. When $\sigma_d=\sqrt{2}\sigma_o$, Eq.~\eqref{eq_pdfen} reduces to known expression for PDF of eigenvalues of GOE ({\it i.e.} $\beta=1$) \cite{Mehta1}, and is given by
\begin{align}
\begin{split}
	\label{eq_pdfeng}
	P(E_1,E_2,...,E_n) &= \frac{1}{\mathcal{Z}_{n,\beta}}\exp\del{-\frac{1}{2}\sum_{i=1}^{n}E_i^2}\prod_{j<k}|E_j-E_k|^\beta\qquad\Bigg( \mathcal{Z}_{n,\beta}=(2\pi)^{n/2}\prod_{j=1}^{n}\frac{\Gamma(1+j\beta/2)}{\Gamma(1+\beta/2)} \Bigg)
	\end{split}
\end{align}
For $N=3$ the \acrshort{pdf} of \acrshort{nns} is obtained from Eq.~\eqref{eq_pdfen} and assumes the form
\begin{align}
	\label{eq_pdfls3sym}
	\begin{split}
	\prob{s} &= C_0s\gls{bsli0}(c_1s^2)\int_{0}^{\infty}dy\gls{bsli0}(c_1(s+y)^2)\gls{bsli0}(c_1y^2)\exp\del{(2c_1-\frac{1}{3\gls{sdd}^2})(y^2+sy+s^2)}y(s+y),\quad \sbr{c_1 = -\frac{1}{C}\Bigg|\frac{1}{\gls{sdd}^2}-\frac{1}{2\gls{sdo}^2}\Bigg|}.
	\end{split}
\end{align}
The above expression needs to be normalized and unfolded numerically and gives excellent match with corresponding simulated data, with $C\approx{10.5}$ for $\gls{sdd}<\sqrt{2}\gls{sdo}$ and $C\approx{8}$ for $\gls{sdd}>\sqrt{2}\gls{sdo}$ and is shown in Fig.~\ref{sym_goe_3_4}\subref{ls_3_all}. It is important to note that Eq.~\eqref{eq_pdfls3sym} does not have the symmetry described in Eq.~\eqref{eq_sym2symmetry}.

In the limit $\sigma_d\gg\sigma_o$, as argued for $N=2$ case, the \acrshort{pdf} of the eigenvalues can be approximated by the \acrshort{pdf} of the matrix and an explicit calculation for $N=3$ gives
\begin{align}
	\label{eq_pdfpsn3s}
	\probcluster{s} &= \frac{9}{4\pi}\exp\del{-\frac{9s^2}{16\pi}}\ferfc{\frac{\sqrt{3}s}{4\sqrt{\pi}}}.
\end{align}
When $\sigma_d=\sqrt{2}\sigma_o$, the PDF of NNS is exactly calculated from Eq.\eqref{eq_pdfls3sym} and us given by
\begin{align}
	\label{eq_pdfgoe3s}
	\probgoe{s} &= \frac{243}{32\pi^2}s^2\exp(-\frac{9}{4\pi}s^2)-\frac{27}{32\pi}s\del{\frac{27}{4\pi}s^2-6}\exp(-\frac{27}{16\pi}s^2)Erfc\del{\frac{3}{4\sqrt{\pi}}s}.
\end{align}
In the limit, $\sigma_d\ll\sigma_o$, we have not been able to find an exact analytical expression but the numerical data suggests that the corresponding NNS distribution is very close to that of GOE. 
Hence, for intermediate statistics, we can again construct a Brody like distribution similar to Eq.~\eqref{eq_phase_2s}. Limiting cases of level clustering, $\prob{s}\sim\exp\del{-3s^2}Erfc\del{s}$ and level repulsion, $\prob{s}\sim\frac{2}{\sqrt{\pi}}s^2\exp\del{-4s^2}-s(2s^2-1)\exp\del{-3s^2}Erfc(s)$ suggests a transition function of the form $\prob{s}\sim\frac{2\eta}{\sqrt{\pi}}s^2\exp\del{-4s^2}-s^\eta(2\eta{}s^{2}-1)\exp\del{-3s^2}Erfc(s)$. Normalized and unfolded the interpolating function is given in appendix (Eq.\eqref{eq_sym3phase1}) and the phase diagram obtained from the estimation of $\eta$ is shown in Fig.~\ref{sym_goe_3_4}\subref{pdiag_ls_00003}.

\captionsetup[subfigure]{position=top, labelfont=bf,textfont=normalfont,singlelinecheck=off,justification=raggedright}
\renewcommand{\x}{0.32}
\begin{figure}[!t]
	\vspace{-10pt}
	\centering
	\subfloat[][]{\includegraphics[width=\x\textwidth, trim=120 0 140 10, clip=true]{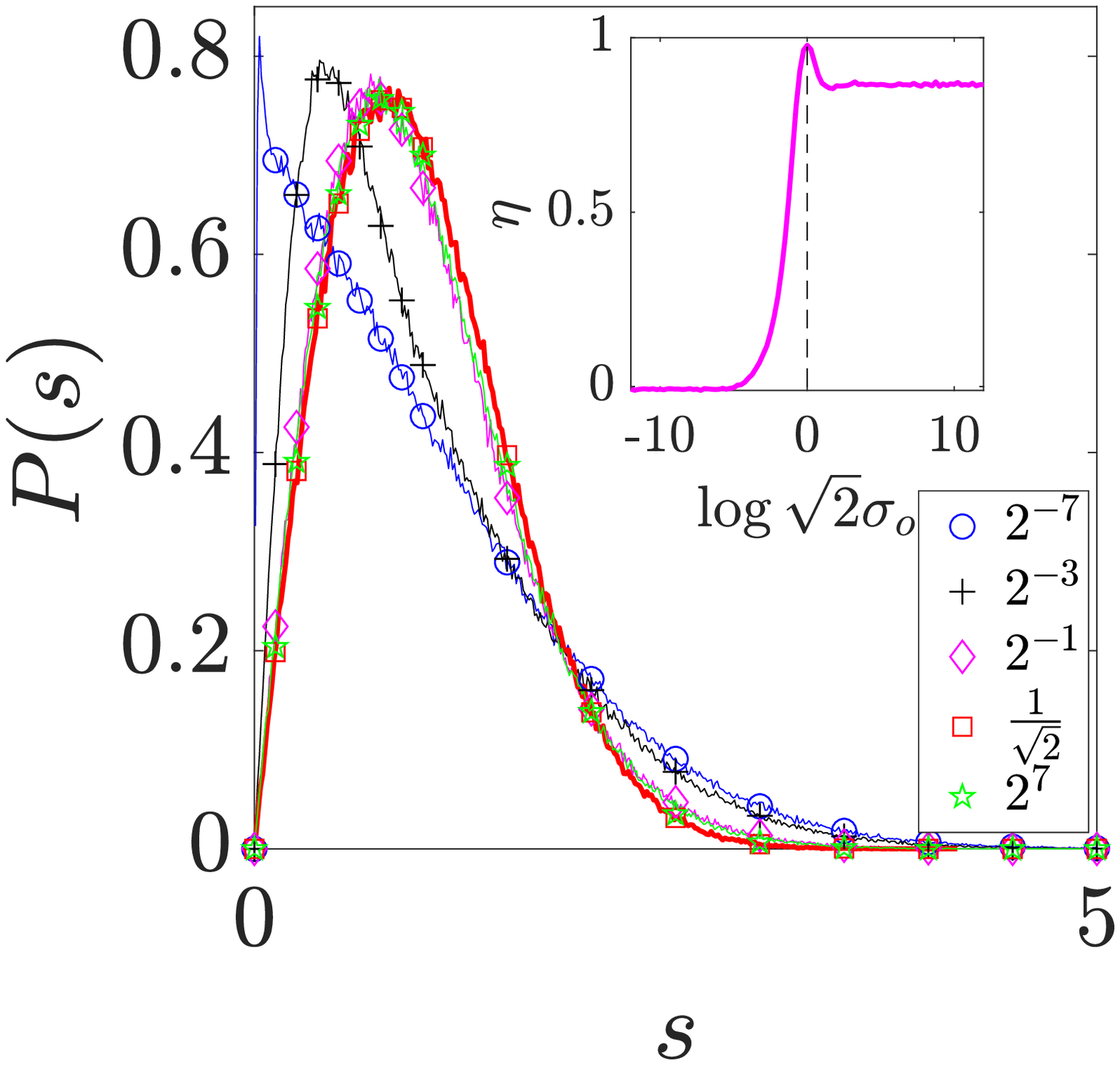}\label{ls_3_all}}
	\subfloat[][]{\includegraphics[width=\x\textwidth, trim=100 0 140 10, clip=true]{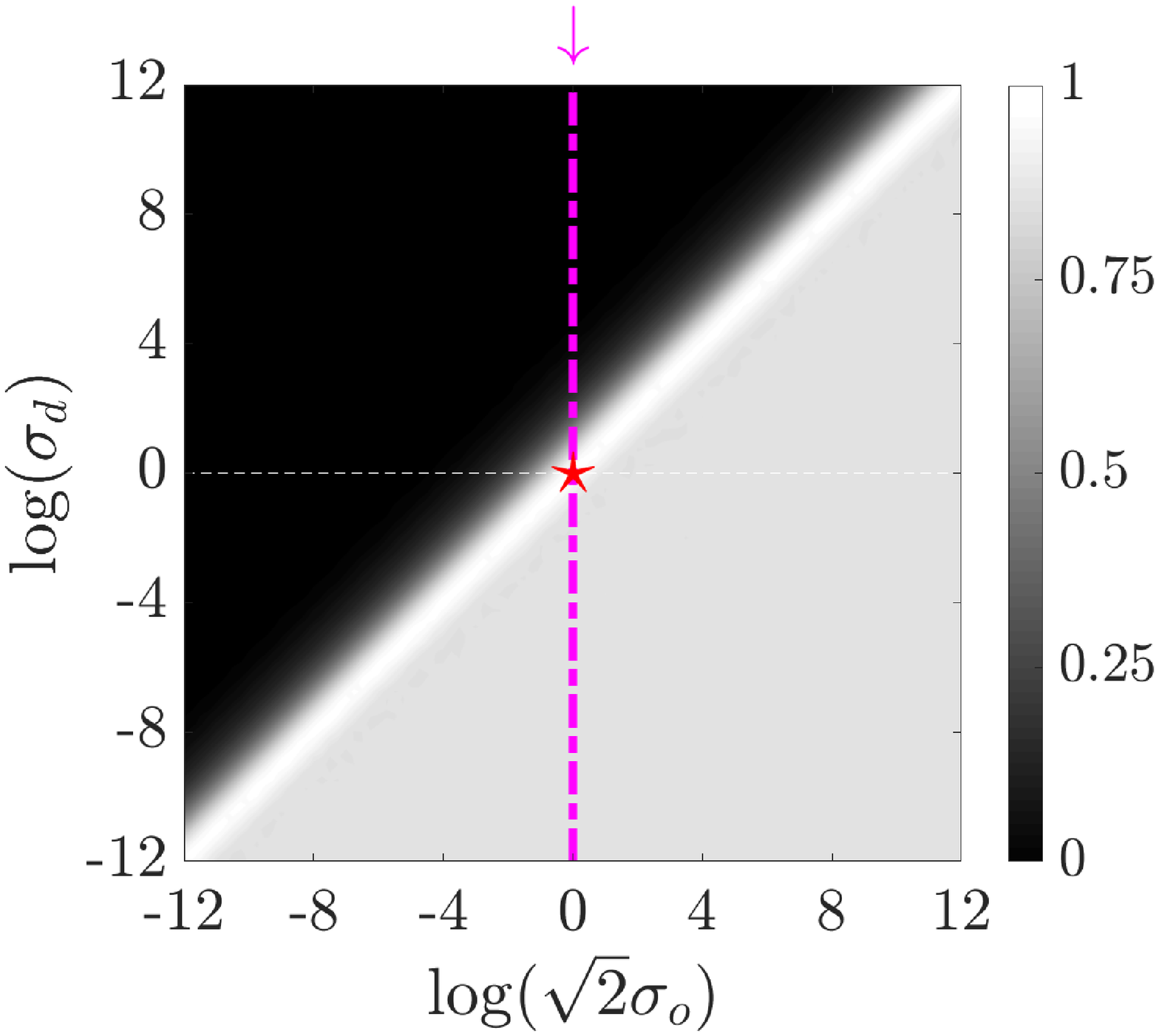}\label{pdiag_ls_00003}}
	\vfill\vspace{-10pt}
	\subfloat[][]{\includegraphics[width=\x\textwidth, trim=100 0 140 10, clip=true]{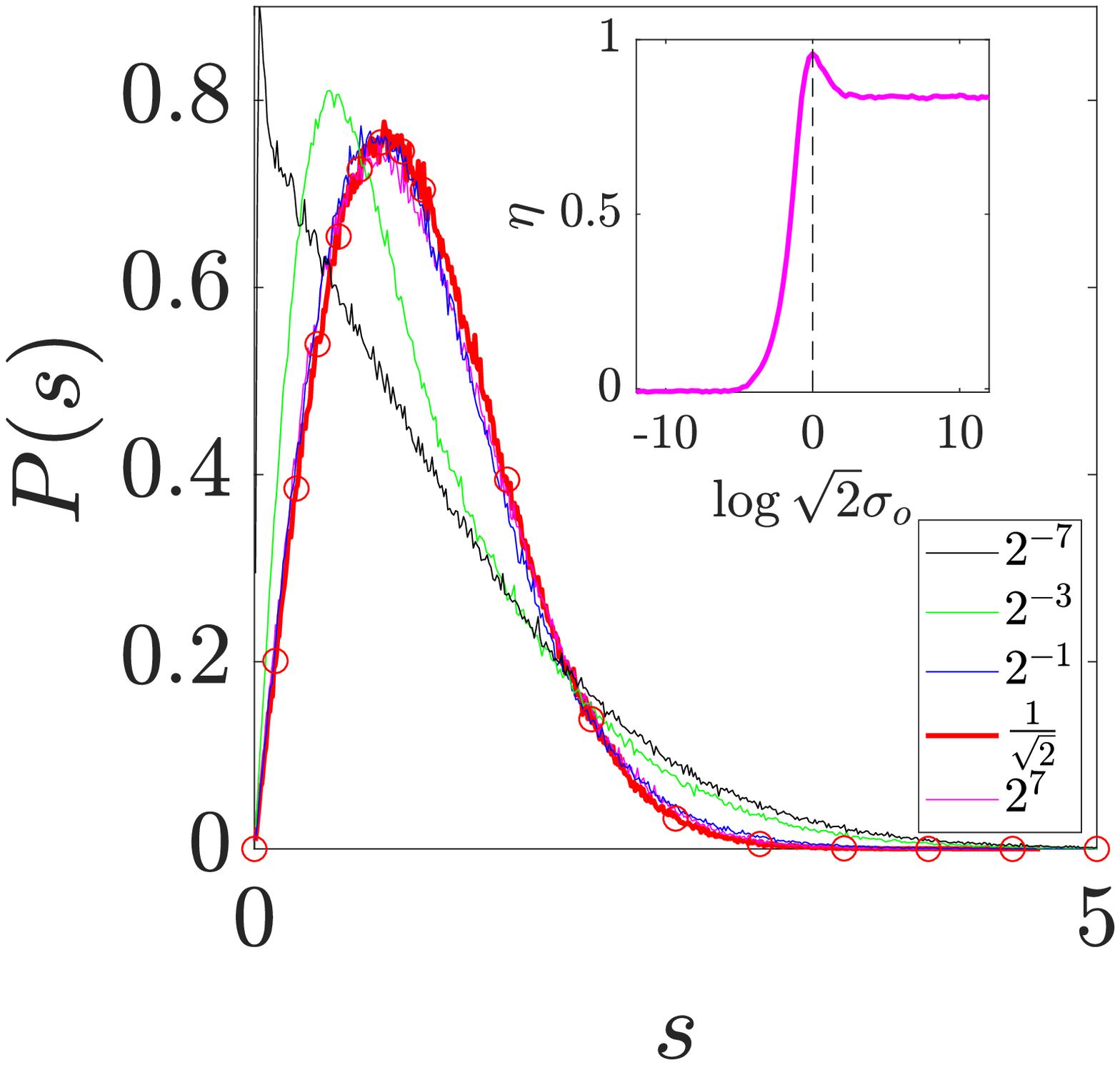}\label{ls_4_all}}
	\subfloat[][]{\includegraphics[width=\x\textwidth, trim=100 0 140 10, clip=true]{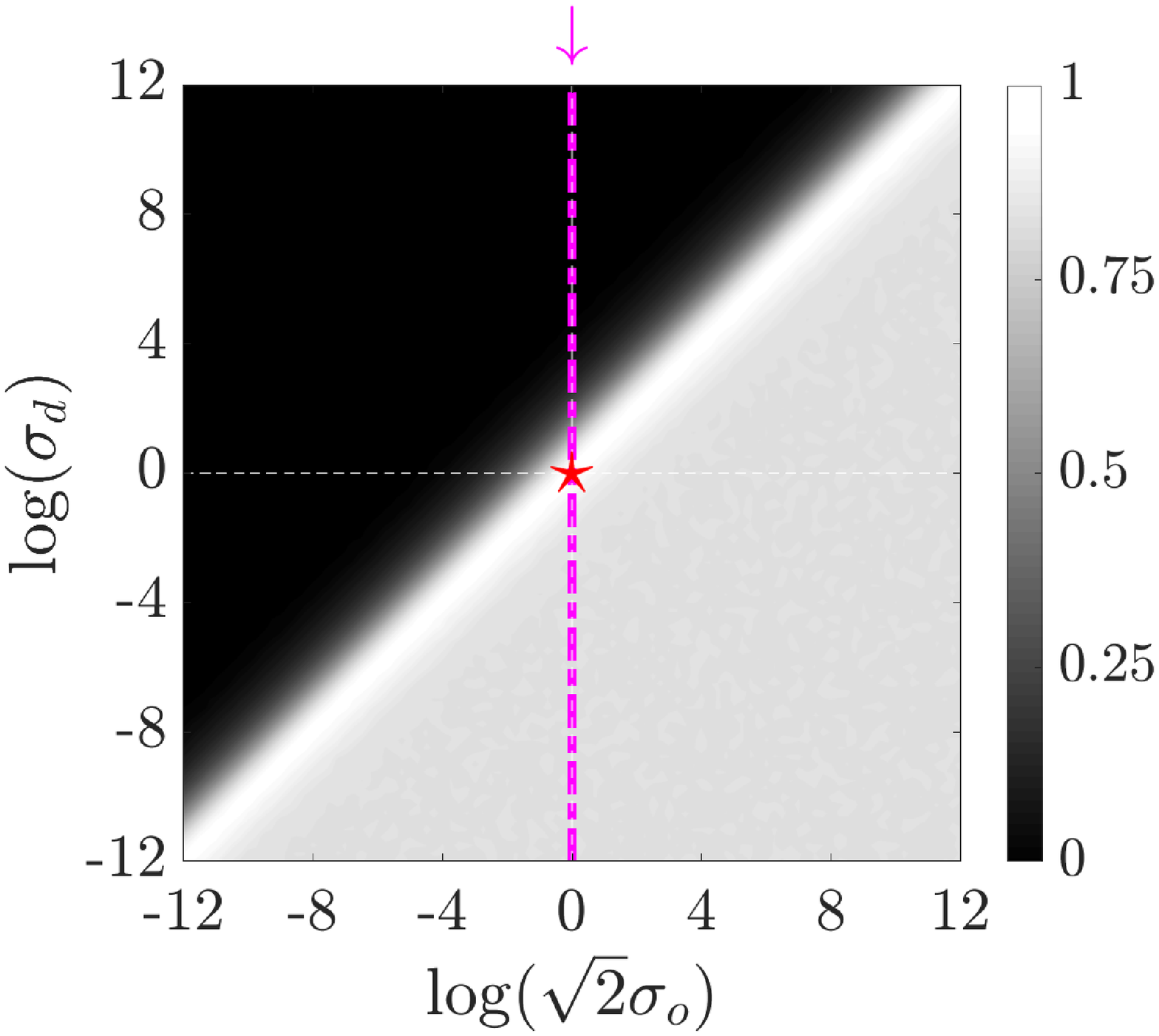}\label{pdiag_ls_00004}}
	\caption[]{{\bf Transition from level clustering to repulsion in NNS of symmetric matrices}: {\acrshort{pdf} of \acrshort{nns} for different $\gls{sdo}$
			with $\gls{sdd}=1$ for \subref{ls_3_all}~{$N=3$} \subref{ls_4_all}~{$N=4$}. Markers in \subref{ls_3_all} denote enumeration of Eq.~\eqref{eq_pdfls3sym}.} 
		Insets in \subref{ls_3_all} and \subref{ls_4_all} show fitted transition parameter $\eta$ vs $\log(\gls{sdd})$ when $\gls{sdo}=1/\sqrt{2}$. Corresponding regions are denoted by bold lines in the \acrshort{pd}. The star in PD indicate GOE.
		{\acrshort{pd} constructed w.r.t. $\eta$ values} \subref{pdiag_ls_00003} $N=3$, \subref{pdiag_ls_00004} $N=4$.}
	\label{sym_goe_3_4}
\end{figure}

\subsection{NNS distributions for $N = 4$}
From the ansatz in Eq.~\eqref{eq_pdfen}, for $N=4$, we obtain the spacing distribution
\begin{align}
	\label{eq_pdfls4sym}
	\begin{split}
	\prob{s} &= C_0sI_0(c_1s^2)\int_{0}^{\infty}dy\int_{0}^{\infty}dz\quad yz(s+y)(y+z)(s+y+z)I_0(c_1y^2)I_0(c_1z^2)I_0(c_1(s+y)^2)I_0(c_1(y+z)^2)\\&\hspace{100pt}I_0(c_1(s+y+z)^2)\exp\del{\del{c_1-\frac{1}{8\sigma_d^2}}\del{3s^2+4y^2+3z^2+2sz+4y(s+z)}}.
	\end{split}
\end{align}
Even numerical evaluation of the above integral is difficult for any $\cbr{\sigma_d,\sigma_o}$ but the GOE limit, $\sigma_d=\sqrt{2}\sigma_o$, can be exactly calculated
\begin{align}
	\label{eq_pdfgoe4s}
	\begin{split}
	\probgoe{s} &= \frac{3\pi}{131072\sqrt{2}} \exp(-\frac{27\pi}{128}s^2) s \Bigg(48 s (448 - 27 \pi s^2) + 9 \sqrt{2} \exp(\frac{9\pi s^2}{128}) (3 \pi s^2 - 64) (9 \pi s^2 - 64) Erfc\del{\frac{3\sqrt{\pi}}{8\sqrt{2}} s} \\&\qquad\qquad+ 512 \sqrt{6} \exp(\frac{3\pi}{128}s^2) (3\pi s^2 - 16) Erfc\del{\frac{1}{8}\sqrt{\frac{3\pi}{2}} s}\Bigg).
	\end{split}
\end{align}
In the clustering limit $\sigma_d\gg\sigma_o$, normalized and unfolded expression is also exactly calculated
\begin{align}
	\label{eq_pdfpsn4s1}
	\probcluster{s} &= \frac{\mu}{C}\del{\sqrt{\pi}\exp\del{-\frac{\mu^2}{4}s^2}Erfc\del{\frac{\mu}{2\sqrt{3}}s}-2\exp\del{-\frac{3\mu^2}{8}s^2}\sum_{j=0}^{\infty}\frac{H_{2j}\del{\frac{\mu}{2\sqrt{2}}s}}{8^{\del{j+\frac{1}{2}}}\fgamma{j+\frac{3}{2}}}}
\end{align}
where $H_j(x)$ is the $j^{th}$ Hermite polynomial [details in Appendix Eq.\eqref{eq_pdfpsn4s},\eqref{eq_erfnormal}]. Normalization constant, $C\approx 1.047198$ and mean of the normalized expression $\mu\approx 0.732364$ are obtained numerically.
Here also in the limit $\sigma_d\ll\sigma_o$, simulations of PDF of NNS  resembles that of GOE. 
For intermediate statistics, constructing a Brody like interpolating function is difficult as the limiting expressions are very complicated. Thus, we can resort to additive RMT approach \cite{Schweiner1} to form a crossover function like
\begin{align}
	\label{eq_crossover}
	\prob{\eta;s}=\eta\probgoe{s}+(1-\eta)\probcluster{s}
\end{align}
from which we estimate $\eta$ to quantify level statistics for any values of $\cbr{\sigma_d,\sigma_o}$
and the results are shown in Fig.~\ref{sym_goe_3_4}\subref{ls_4_all},\subref{pdiag_ls_00003}.

\subsection{ RNNS distributions for $N = 3$}

 \captionsetup[subfigure]{position=top, labelfont=bf,textfont=normalfont,singlelinecheck=off,justification=raggedright}
\renewcommand{\x}{0.32}
\begin{figure}[!t]
	\vspace{-10pt}
	\centering
	\subfloat[][]{\includegraphics[width=\x\textwidth, trim=120 0 140 10, clip=true]{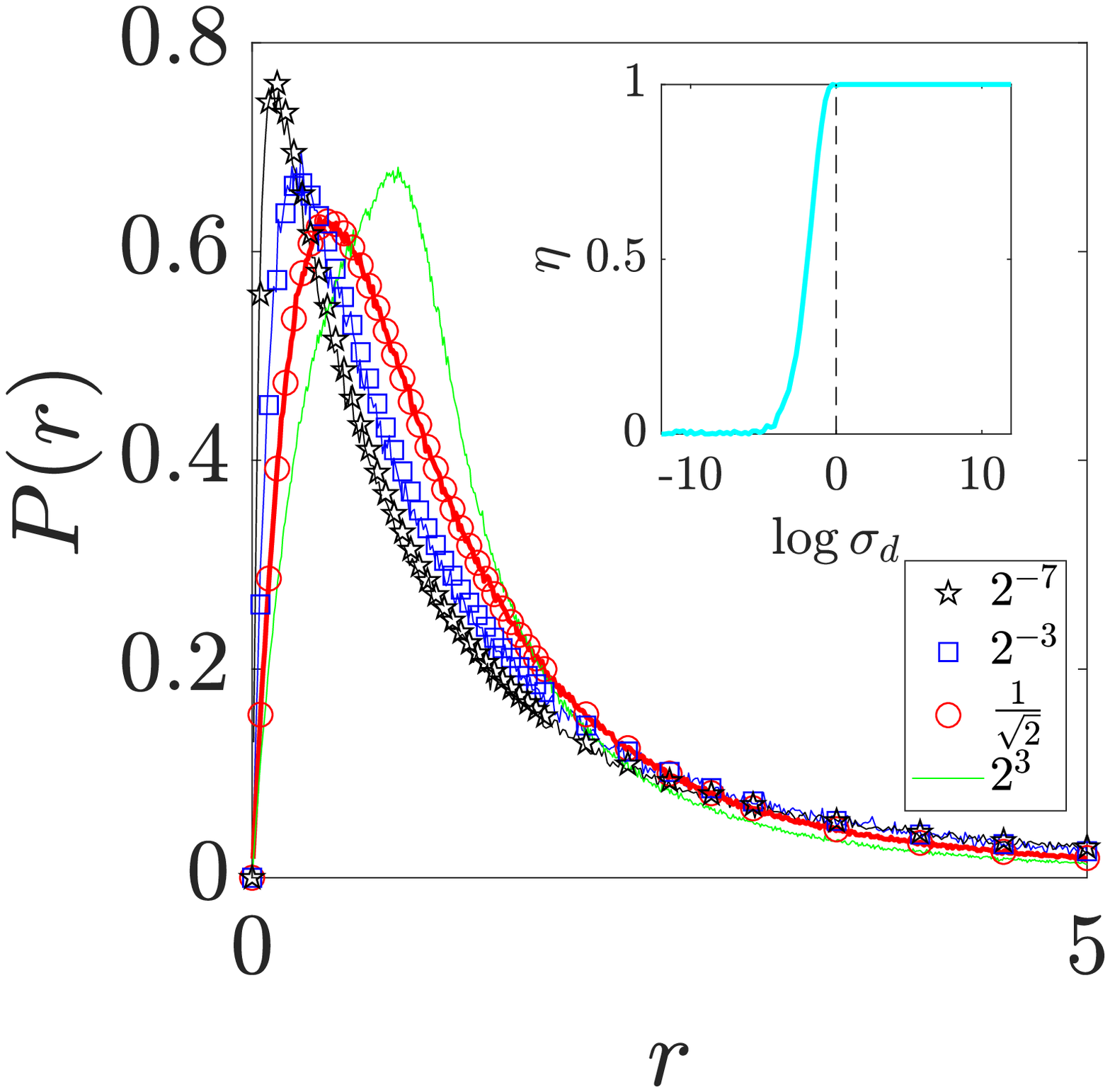}\label{rl_3_all}}
	\subfloat[][]{\includegraphics[width=\x\textwidth, trim=100 0 140 10, clip=true]{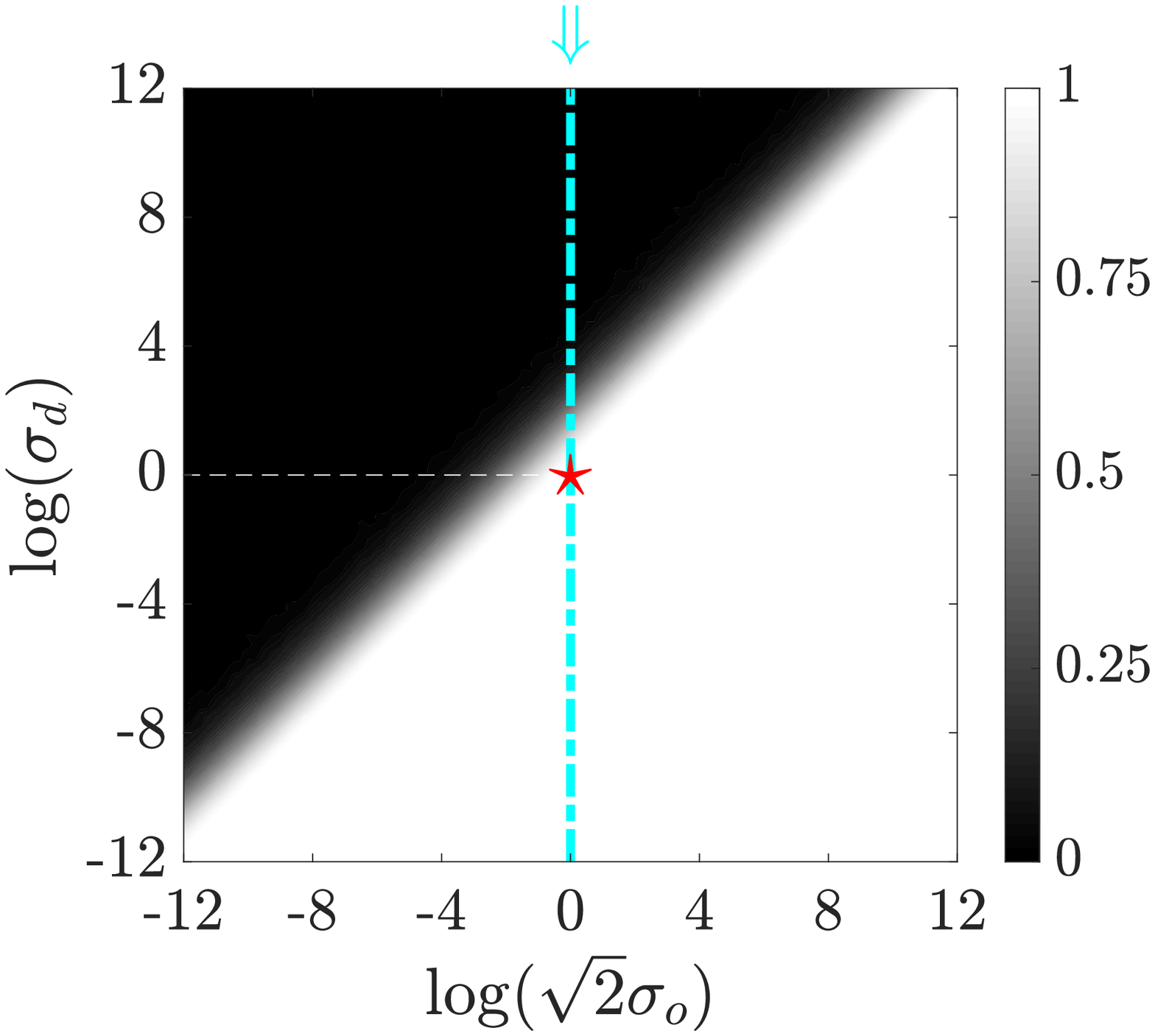}\label{pdiag_rl_00003}}
	\vfill\vspace{-10pt}
	\subfloat[][]{\includegraphics[width=\x\textwidth, trim=120 0 140 10, clip=true]{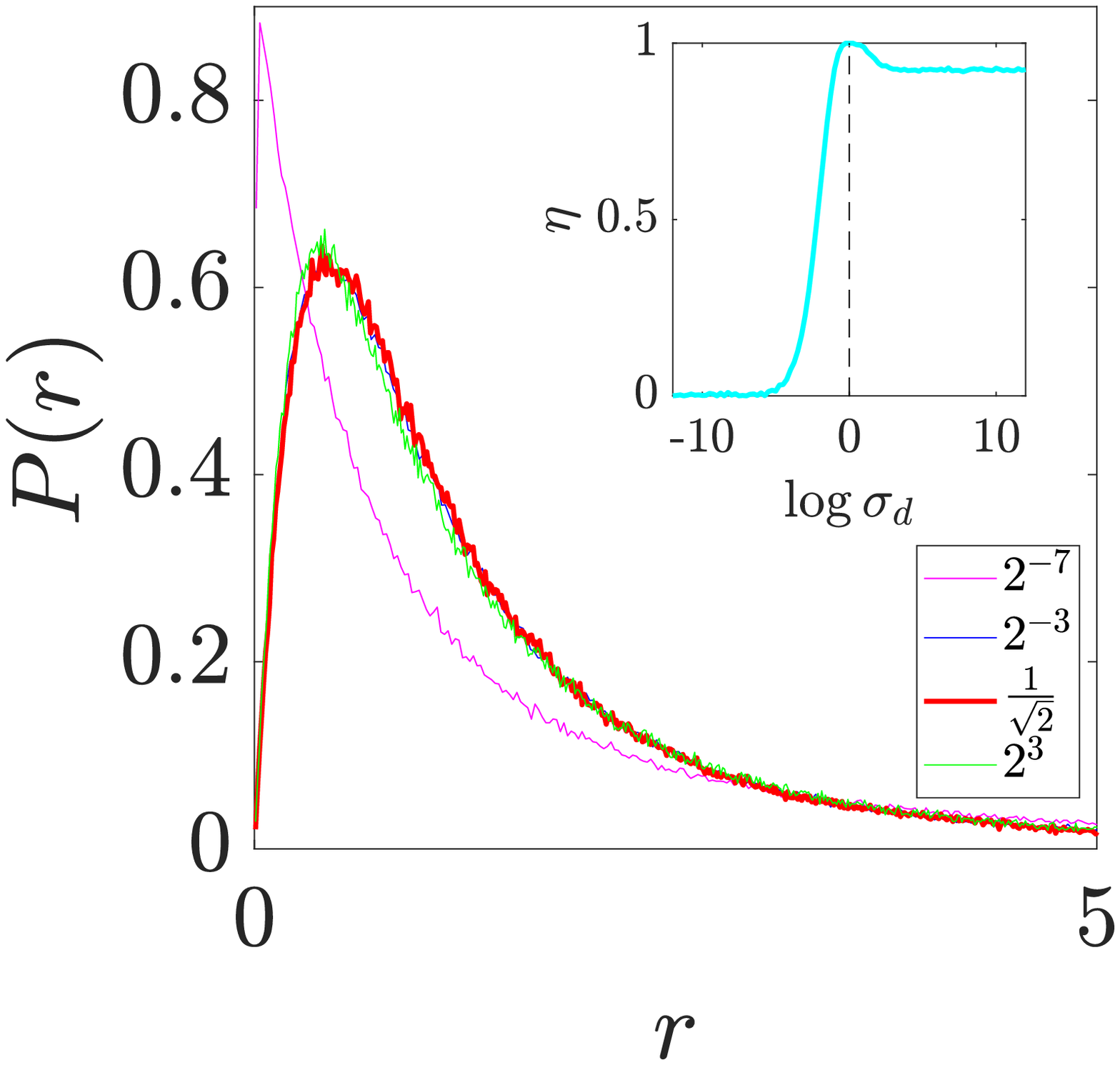}\label{rl_4_all}}
	\subfloat[][]{\includegraphics[width=\x\textwidth, trim=100 0 140 10, clip=true]{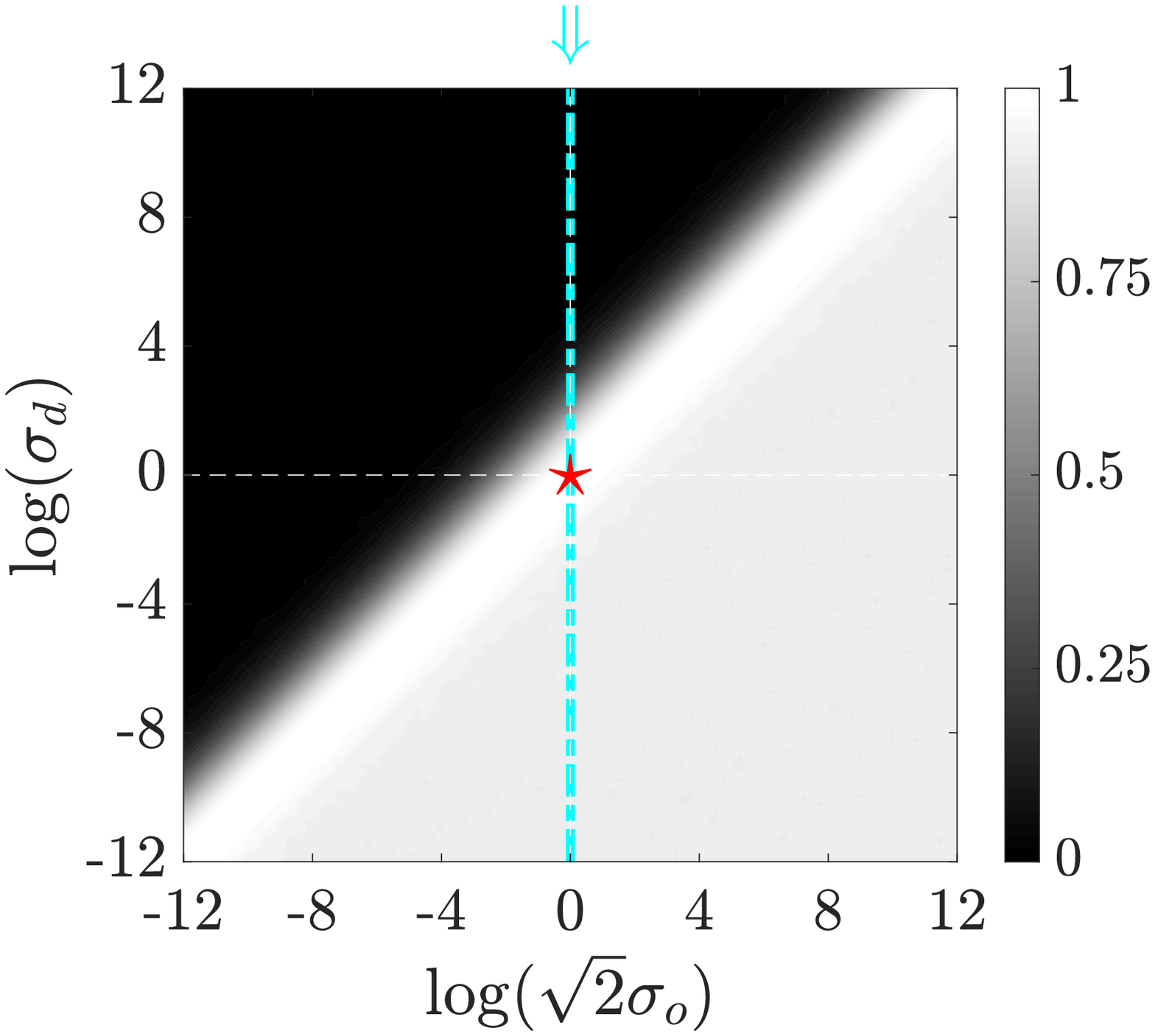}\label{pdiag_rl_00004}}
		\caption[]{{\bf Transition from level clustering to repulsion in RNNS of symmetric matrices}:
	{\acrshort{pdf} of \acrshort{rnns} for different $\gls{sdo}$
		with $\gls{sdd}=1$ for \subref{rl_3_all}~{$N=3$} \subref{rl_4_all}~{$N=4$}. Markers in \subref{rl_3_all} denote enumeration of Eq.~\eqref{eq_pdfrl3sym}.} 
	Insets in \subref{rl_3_all} and \subref{rl_4_all} show fitted transition parameter $\eta$ vs $\log(\gls{sdd})$ when $\gls{sdo}=1/\sqrt{2}$. Corresponding regions are denoted by bold lines in the \acrshort{pd}. The star in PD indicate GOE.
	{\acrshort{pd} constructed w.r.t. $\eta$ values} \subref{pdiag_rl_00003} $N=3$, \subref{pdiag_rl_00004} $N=4$.}
	\label{sym_goe_3r}
\end{figure}

The numerical issues in unfolding \acrshort{nns} distributions can be avoided by estimating distribution of ratio of level spacings among consecutive eigenvalues (\acrshort{rnns}) defined as
\begin{equation}
\tilde{r}_i = \frac{\min(s_i,s_{i-1})}{\max(s_i,s_{i-1})} = \min\left(r_i, \frac{1}{r_i} \right)\nonumber
\end{equation}
where $r_i = s_i/s_{i-1}$ \cite{Oganesyan1}.
 \acrshort{rnns} can be defined for matrices with size, $N\geq{3}$, and for $N=3$ \acrshort{pdf} of \acrshort{rnns} is 
\begin{align}
	\label{eq_pdfr}
	\prob{r} = \int_{-\infty}^{\infty}d\gls{E2}\int_{-\infty}^{\gls{E2}}d\gls{E1}\int_{\gls{E2}}^{\infty}d\gls{E3}\prob{\mathcal{E}}\delta\del{r-\frac{\gls{E3}-\gls{E2}}{\gls{E2}-\gls{E1}}}.
\end{align}
Once again using $\prob{\mathcal{E}}$ from Eq.~\eqref{eq_pdfen} we can get \acrshort{pdf} of \acrshort{rnns} 
\begin{align}
	\label{eq_pdfrl3sym}
	\begin{split}	
	\prob{r} &= C_0(r+r^2)\int_{0}^{\infty}dx\Bigg[x^4\gls{bsli0}(c_1(1+r)^2x^2)\gls{bsli0}(c_1r^2x^2)\gls{bsli0}(c_1x^2)\exp\del{\del{2c_1-\frac{1}{3\gls{sdd}^2}}(1+r+r^2)x^2}\Bigg].
	\end{split}
\end{align}
As \acrshort{rnns} is the ratio of two consecutive levels, in the level clustering regime $\prob{r}$ does not require any local transformation to bring out the universal features. For N=3 we can calculate the RNNS distribution as
\begin{align}
	\label{eq_pdfpsn3r}
	\probcluster{r} &= \frac{3\sqrt{3}}{2\pi}\frac{1}{1+r+r^2}.
\end{align} 
Again in the GOE limit, $\gls{sdd}=\sqrt{2}\gls{sdo}$ the \acrshort{pdf} of \acrshort{rnns} has also been obtained  \cite{Atas1}
\begin{align}
	\label{eq_pdfgoer3}
	\probgoe{r} &= \frac{27}{8}\frac{r+r^2}{(1+r+r^2)^{-5/2}}.
\end{align}
However, in the limit $\sigma_d\ll\sigma_o$, numerical evaluation of Eq.~\eqref{eq_pdfrl3sym} shows deviation from simulated data. We observe that as $\sigma_o$ becomes larger than $\sigma_d$, level repulsion is present, but the functional form starts deviating from that of GOE in contrast to our findings for NNS distributions. For intermediate statistics we can construct an interpolating function between $P_{cluster}(s)$ and $P_{GOE}(s)$ but the region $\sigma_d\ll\sigma_o$ will not be correctly represented. Instead we define an additive crossover function for this purpose as in Eq.~\eqref{eq_crossover}
and estimate $\eta$ to represent the phase diagram in Fig.\ref{sym_goe_3r}\subref{pdiag_rl_00003}.

\subsection{ RNNS distributions for $N = 4$}

The PDF of RNNS can be obtained for the clustering regime $\sigma_d\gg\sigma_o$
\begin{align}
	\label{eq_pdfpsn4r}
	\probcluster{r} &= \frac{3}{2\pi}\del{2\sqrt{3}-\frac{2+r}{\sqrt{3r^2+4r+4}}-\frac{1+2r}{\sqrt{4r^2+4r+3}}}\frac{1}{1+r+r^2}.
\end{align}
For GOE case, analytical calculation is straightforward but the final expression is very lengthy and has already been reported \cite{Atas2}. Interestingly, in the limit $\sigma_d\ll\sigma_o$, simulations show that RNNS distributions have stronger similarity to that of GOE in comparison to deviations observed behaviour for $N=3$. 
Here also we use an interpolating function as in Eq.~\eqref{eq_crossover} and represent the phase diagram in terms of estimated $\eta$ in Fig.\ref{sym_goe_3r}\subref{pdiag_rl_00004}.

\section[Hermitian matrices]{Hermitian matrices}{\label{sect_her}}

Let, $\gls{H}$ be a $N{\times}N$ Hermitian matrix, $\ni \gls{Hij}=\gls{Hji}^*$.
The diagonal elements are such that
$\gls{Hii}{\sim}{\cal N}(0,\gls{sdd}^2)$ and the offdiagonal elements $\gls{Hij}=\gls{xij}+i\gls{yij}$ with $\quad\gls{xij},\gls{yij}{\sim}{\cal N}(0,\gls{sdo}^2)$ ${\quad\forall\quad}i{\neq}j$. Using Eq.~\eqref{eq_fullpdf}, we get \acrshort{pdf} of such matrices
\begin{align}
	\label{eq_pdfherh}
	\begin{split}
	\prob{\gls{H}} &= C\exp\del{-\sum_{i=1}^{N}\frac{\gls{Hii}^2}{2\gls{sdd}^2}-\sum_{i<j}^{N}\frac{\gls{xij}^2+\gls{yij}^2}{2\gls{sdo}^2}}\\
	&= C\exp\del{\del{\frac{1}{4\gls{sdo}^2}-\frac{1}{2\gls{sdd}^2}}\sum_{i=1}^{N}\gls{Hii}^2-\frac{\text{Tr}(\gls{H}^2)}{4\gls{sdo}^2}}\\
	&\sbr{\text{where, }C=\frac{(2\pi)^{\frac{N^2}{2}}}{\gls{sdd}^N\gls{sdo}^{N(N-1)}}}.
	\end{split}
\end{align}
We can use the similarity transformation, $\gls{H}=\gls{U}^\dagger\mathcal{E}\gls{U}$, by which we get $\text{Tr}(\gls{H}^2) = \sum_{i}^{N}\gls{Ei}^2$. But here going from matrix space to eigenspace requires the Jacobian $J\del{H\to\cbr{\mathcal{E},U}}=\prod_{j<k}^{N}\del{\gls{Ei}-\gls{Ej}}^2$ \cite{Livan1}. Using this in Eq.~\eqref{eq_pdfherh}, then following Eq.~\eqref{eq_pdfhtoe}, \eqref{eq_pdfEtoe} we get,
\begin{align}
	\label{eq_hermainpdfe}
	&\prob{\mathbf{E}}\propto \exp\del{-\frac{1}{4\gls{sdo}^2}\sum_{i}^{N}\gls{Ei}^2}\prod_{i<j}^{N}\del{\gls{Ei}-\gls{Ej}}^2
\end{align}

\captionsetup[subfigure]{position=top, labelfont=bf,textfont=normalfont,singlelinecheck=off,justification=raggedright}
\renewcommand{\x}{0.33}
\begin{figure}[!t]
	\vspace{-10pt}
	\centering
	\subfloat[][]{\includegraphics[width=\x\textwidth, trim=90 0 100 10, clip=true]{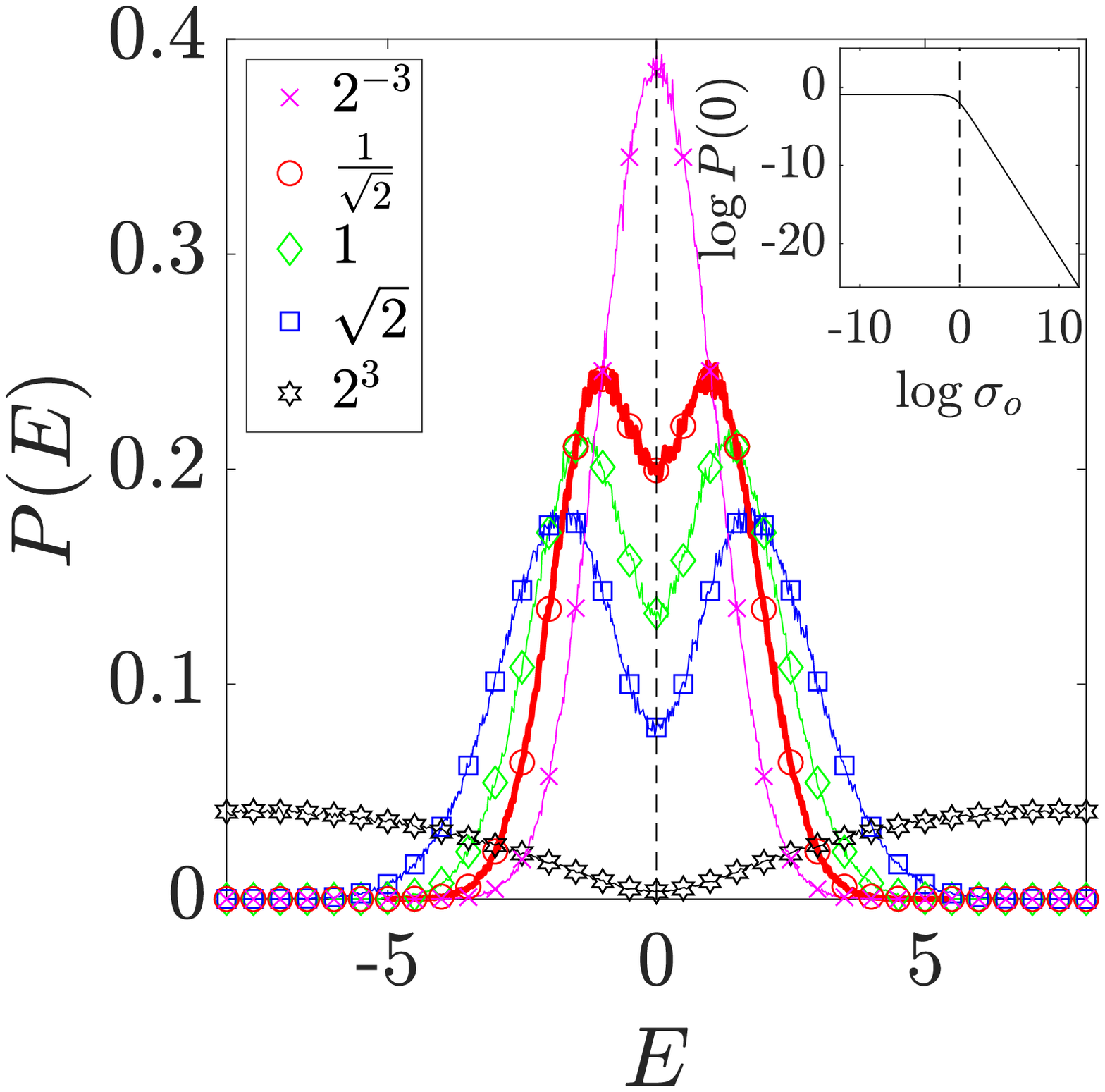}\label{eig_her_all}}
	\subfloat[][]{\includegraphics[width=\x\textwidth, trim=90 0 100 10, clip=true]{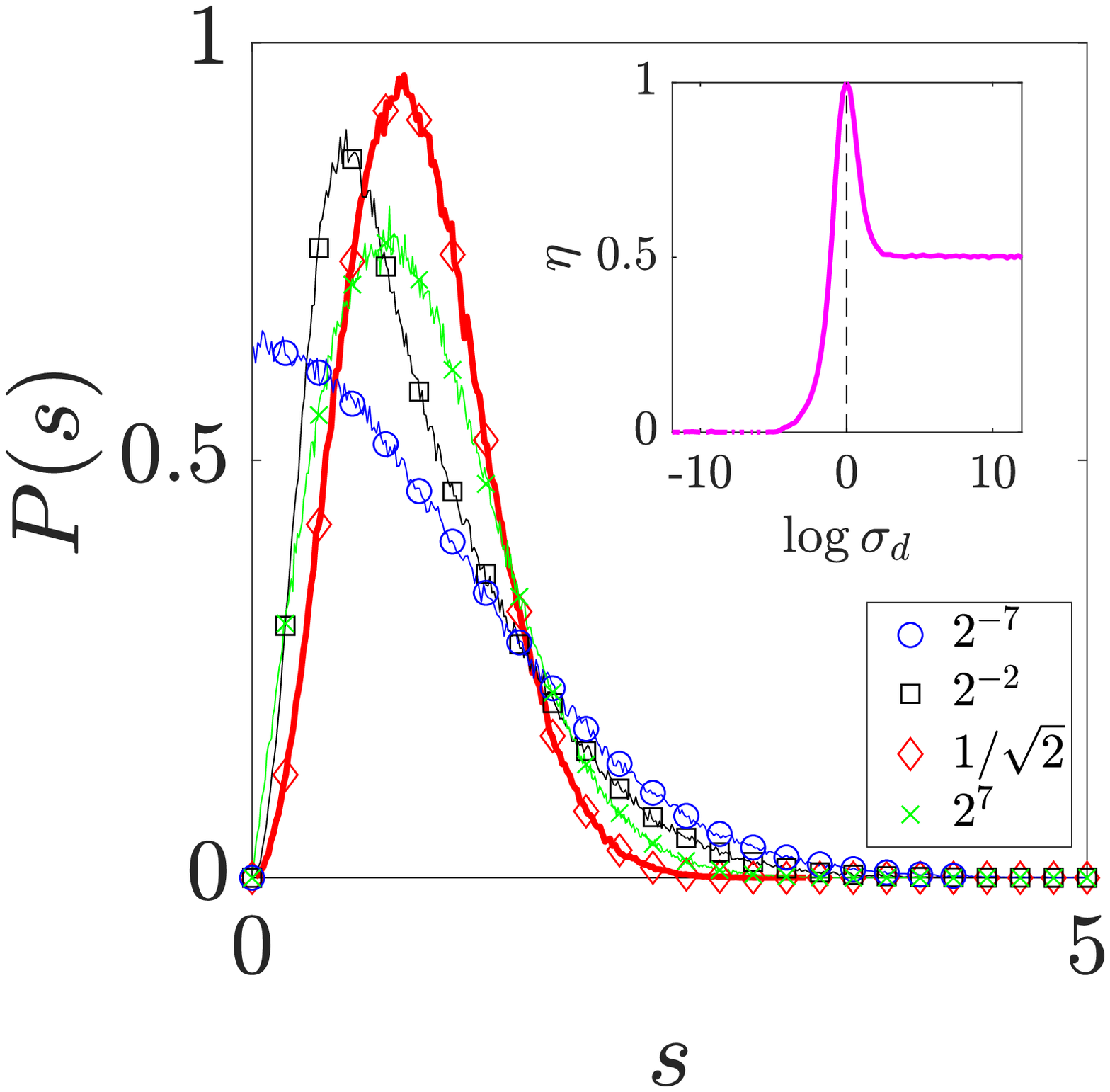}\label{ls_her_all}}
	\subfloat[][]{\includegraphics[width=\x\textwidth, trim=90 0 100 10, clip=true]{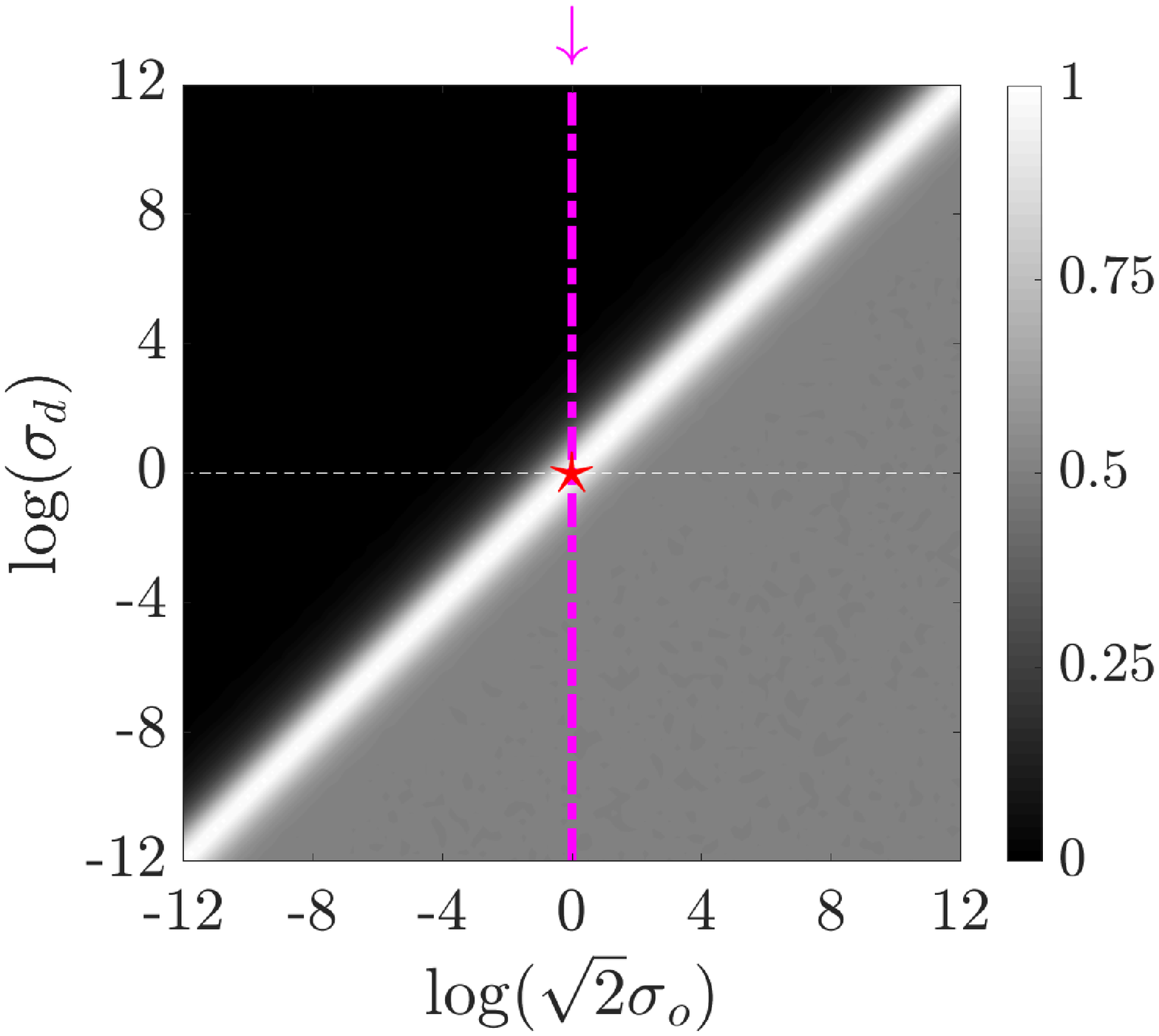}\label{pdiag_her_2}}
	\caption[]{{\bf Results for $2\times2$ Hermitian matrix}:
		\subref{eig_her_all} PDF of eigenvalue for different $\gls{sdo}$ with $\gls{sdd}=1$. 
		Markers denote expression in Eq.~\eqref{eq_marginalher} and continuous lines are simulations.
		The red line corresponds to GUE.
		Inset shows $\prob{E=0}$ vs \gls{sdo} in log-log scale. 
		\subref{ls_her_all} \acrshort{pdf} of  ~{\acrshort{nns}}. 
		Inset shows transition parameter $\eta$ vs $\log(\gls{sdd})$ when $\gls{sdo}=1/\sqrt{2}$. 
		\subref{pdiag_her_2}~Phase diagram of crossover from level clustering (dark) to level repulsion (light).  
		The red star indicates GUE obtained from Eq.~\eqref{eq_herpdfs2}. 
		The inset in \subref{ls_all} corresponds to the dot-dashed line marked with arrow.
		In simulations averages are computed over 500000 samples.}
	\label{pdiag_her}
\end{figure}

\subsection{NNS distributions for $N = 2$}
For $N=2$ we can take an arbitrary unitary matrix, i.e. $U=e^{i\alpha}\begin{pmatrix}
	\cos\theta{e^{-i\phi}}&\sin\theta\\-\sin\theta&\cos\theta{e}^{i\phi}
\end{pmatrix}$ giving us
\begin{align}
	\label{eq_trans2her}
	\begin{bmatrix}
		\gls{H11}&x_{12}+iy_{12}\\x_{12}-iy_{12}&\gls{H22}
	\end{bmatrix} = \begin{bmatrix}
		\gls{E1}\cos^2\theta+\gls{E2}\sin^2\theta&(\gls{E1}-\gls{E2})\sin\theta\cos\theta(\cos\phi+i\sin\phi)\\
		(\gls{E1}-\gls{E2})\sin\theta\cos\theta(\cos\phi-i\sin\phi)&\gls{E2}\cos^2\theta+\gls{E1}\sin^2\theta
	\end{bmatrix}.
\end{align}
We can see that transformation rules for diagonal elements are same as those of the symmetric matrix. 
Substituting Eq.~\eqref{eq_trans2her} in Eq.~\eqref{eq_pdfherh} and using Eq.~\eqref{eq_pdfhtoe} along with normalization we get
\begin{align}
	\label{eq_herpdfe1e2}
	\begin{split}
	&\prob{\mathbf{E}} = \frac{|\gls{E1}-\gls{E2}|}{4\gls{sdd}\gls{sdo}\sqrt{2\pi|2\gls{sdo}^2-\gls{sdd}|}}\exp\del{-\sum_{i=1}^{2}\frac{\gls{Ei}^2}{2\gls{sdd}^2}}\exp\del{\del{\frac{1}{4\gls{sdd}^2}-\frac{1}{8\gls{sdo}^2}}(\gls{E1}-\gls{E2})^2}g\del{\frac{|\gls{E1}-\gls{E2}|}{2}}\\
	&g(x) = \begin{cases}
	\ferf{\sqrt{\frac{1}{\gls{sdd}^2}-\frac{1}{2\gls{sdo}^2}}x},\quad \gls{sdd}<\sqrt{2}\gls{sdo}\\
	\ferfi{\sqrt{\frac{1}{2\gls{sdo}^2}-\frac{1}{\gls{sdd}^2}}x},\quad \gls{sdd}>\sqrt{2}\gls{sdo}
	\end{cases}
	\end{split}
\end{align}
and the corresponding marginal \acrshort{pdf} using Eq.~\eqref{eq_pdfEtoe} can be expressed as
\begin{align}
	\label{eq_marginalher}
	\prob{E} &= \frac{1}{4\sqrt{2\pi}\gls{sdo}\gls{sdd}\sqrt{|2\gls{sdo}^2-\gls{sdd}^2|}}\int_{-\infty}^{\infty}dy|y|\exp\del{-\frac{(y+2E)^2}{4\gls{sdd}^2}-\frac{y^2}{8\gls{sdo}^2}}g(|y|).
\end{align}
The \acrshort{pdf} of \acrshort{nns} is given exactly for generalized $2\times2$ Hermitian matrices
\begin{align}
	\label{eq_herpdfs2}
	\begin{split}
	& \prob{s} = \frac{\mu^2s}{2\sqrt{2}\gls{sdo}\sqrt{|2\gls{sdo}^2-\gls{sdd}^2|}}\exp\del{-\frac{\mu^2s^2}{8\gls{sdo}^2}}g\del{\frac{s}{2}}\\
	& \mu = \begin{cases}
	\frac{2\gls{sdd}}{\sqrt{\pi}}+\frac{4\gls{sdo}^2}{\sqrt{\pi(2\gls{sdo}^2-\gls{sdd}^2)}}\cot^{-1}\del{\frac{\gls{sdd}}{\sqrt{2\gls{sdo}^2-\gls{sdd}^2}}},\quad \gls{sdd}<\sqrt{2}\gls{sdo}\\
	\frac{2\gls{sdd}}{\sqrt{\pi}}+\frac{4\gls{sdo}^2}{\sqrt{\pi(\gls{sdd}^2-2\gls{sdo}^2)}}\tanh^{-1}\del{\frac{\gls{sdd}}{\sqrt{\gls{sdd}^2-2\gls{sdo}^2}}},\quad \gls{sdd}>\sqrt{2}\gls{sdo}
	\end{cases}.
	\end{split}
\end{align}

When $\gls{sdd}\gg\gls{sdo}$, then \gls{H} can be assumed to be a diagonal matrix with all real entries resulting in a similar symmetric matrix of the previous section implying that the $P_{cluster}(s)$ is given by
Eq.(\ref{eq_pdfpoissons}).
However, Eq.~\eqref{eq_herpdfs2} does not have the symmetry of Eq.~\eqref{eq_sym2symmetry}
and existence of clustered states is not evident for $\gls{sdd}\ll\gls{sdo}$.

The PDF in Eq.~\eqref{eq_herpdfe1e2} correctly yields the expected results for \acrshort{gue} when $\gls{sdd}=1,\gls{sdo}=1/\sqrt{2}$ upon using the limits
\begin{align}
	\label{eq_limithereg}
	&\lim\limits_{x\to{0}}\frac{\ferf{x}}{x}=\lim\limits_{x\to{0}}\frac{\ferfi{x}}{x}=\frac{2}{\sqrt{\pi}}.
\end{align}
The \acrshort{jpdf} of eigenvalues is given by
\begin{align}
	\label{eq_pdfe1e2gue}
	\probgue{\gls{E1},\gls{E2}} &= \frac{1}{4\pi}\del{\gls{E1}-\gls{E2}}^2\exp\del{-\frac{\gls{E1}^2+\gls{E2}^2}{2}}
\end{align}
with the corresponding marginal \acrshort{pdf} 
\begin{align}
	\label{eq_pdfegue}
	\probgue{\gls{E}} &= \frac{1+\gls{E}^2}{2\sqrt{2\pi}}\exp\del{-\frac{E^2}{2}}.
\end{align}
From Eq.~\eqref{eq_herpdfs2} we get \acrshort{pdf} of \acrshort{nns} for \acrshort{gue} 
\begin{align}
	\label{eq_pdfs2gue}
	\probgue{s} &= \frac{32}{\pi^2} s^2\exp\del{-4s^2/\pi}
\end{align}
showing quadratic level repulsion {\it i.e.} the familiar form for $\beta=2$ obtained by using the limits
\begin{align}
	\label{eq_limithers}
	\lim\limits_{x\to{0}}\frac{\cot^{-1}(1/x)}{x} &= \lim\limits_{x\to{0}}\frac{\tanh^{-1}x}{x}=1.
\end{align}
Surprisingly in the limit $\gls{sdd}\ll\gls{sdo}$, using $\lim\limits_{x\to\infty}\ferf{x}=1$ in Eq.~\eqref{eq_herpdfs2} we get linear level repulsion as in Eq.~\eqref{eq_pdfs2goe}.
Thus it is possible to access three distinct phases as $\tilde{\sigma}$ is varied. 
This is an improvement over crossover function described in \cite{Schweiner1}, which uses one parameter to break integrability and another one to break anti-unitary symmetry.\\
The limiting cases for level repulsion, $\prob{s}\sim s^2\exp\del{-s^2}$ and level clustering, $\prob{s}\sim\exp\del{-s^2}$, suggest a transition function of the form $\prob{s}\sim{s^{2\eta}}\exp\del{-s^2}$. Normalizing and unfolding this equation we get
\begin{align}
	\label{eq_phase_2s_her}
	\prob{\eta;s} &= \frac{2\fgamma{1+\eta}^{1+2\eta}}{\fgamma{\frac{1}{2}+\eta}^{2(1+\eta)}}s^{2\eta}\exp\del{-\frac{\fgamma{1+\eta}^2}{\fgamma{\frac{1}{2}+\eta}^2}s^2}
\end{align}
Above equation has been used to fit the simulation data and estimate $\eta$ to generate the phase diagram in Fig.\ref{pdiag_her}.

\captionsetup[subfigure]{position=top, labelfont=bf,textfont=normalfont,singlelinecheck=off,justification=raggedright}
\renewcommand{\x}{0.35}
\begin{figure}[!t]
	\vspace{-10pt}
	\centering
	\subfloat[][]{\includegraphics[width=\x\textwidth, trim=120 0 140 10, clip=true]{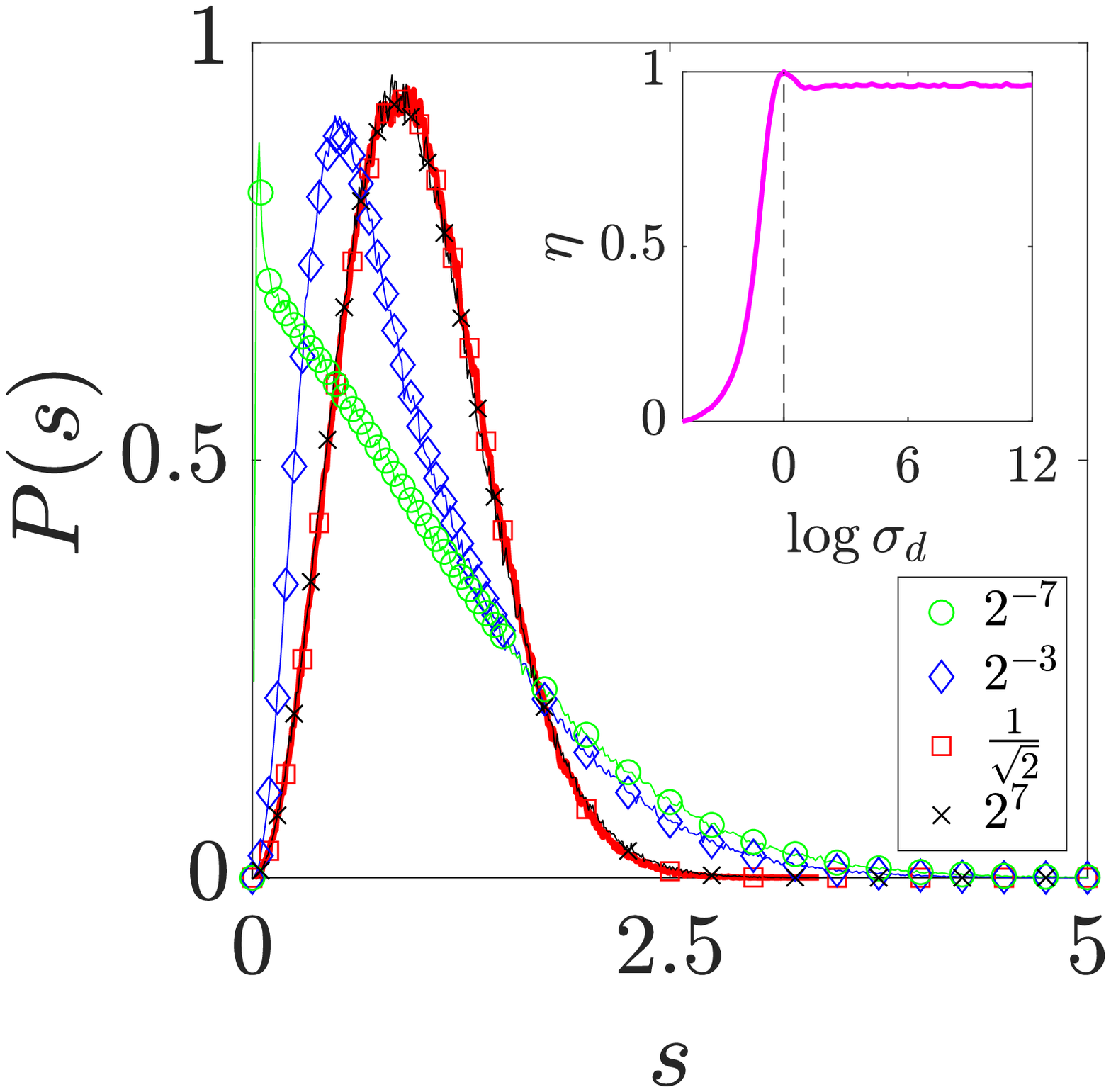}\label{ls_her_3_all}}
	\subfloat[][]{\includegraphics[width=\x\textwidth, trim=100 0 140 10, clip=true]{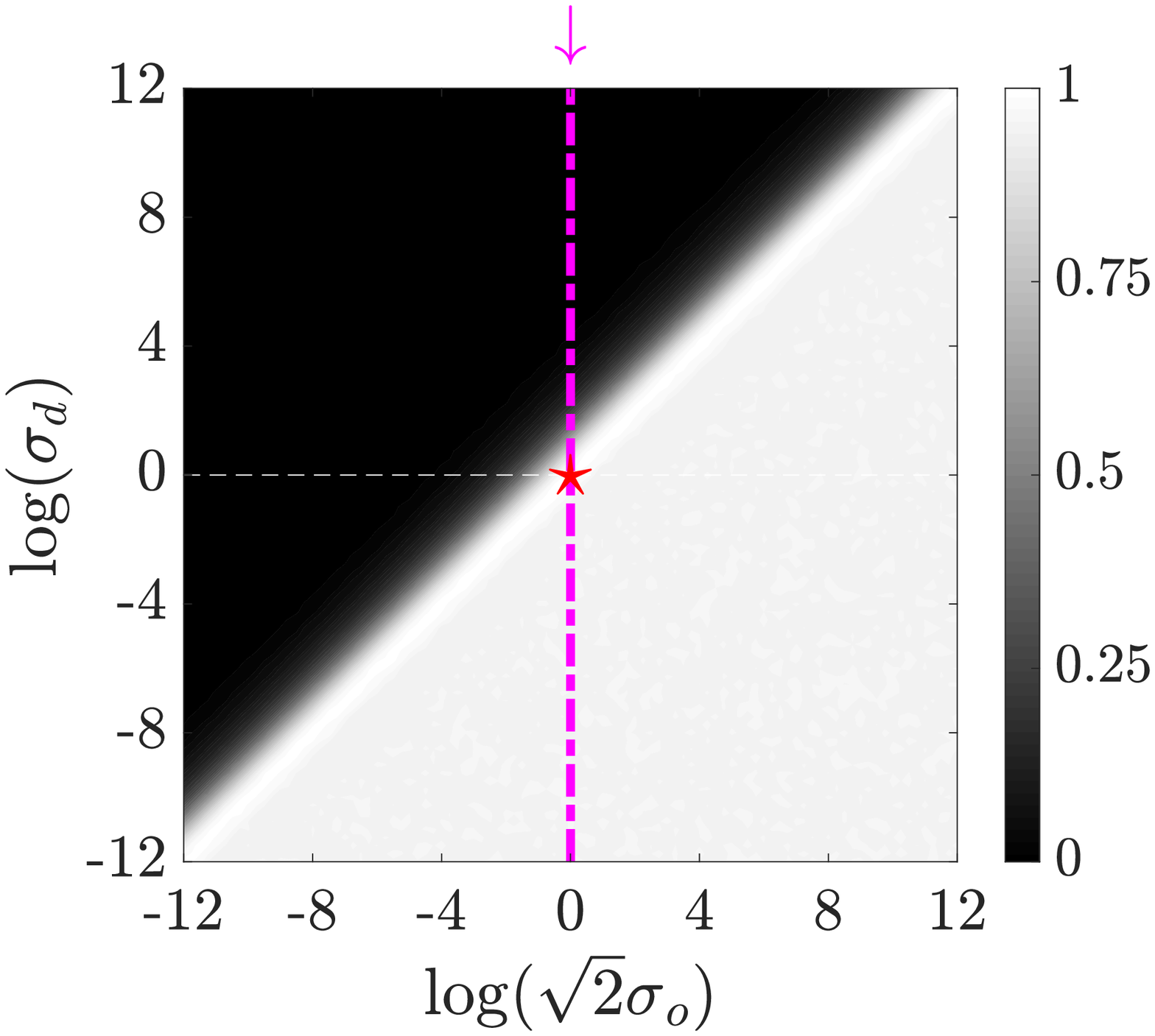}\label{pdiag_her_ls_3}}
	\vfill
	\vspace{-10pt}
	\subfloat[][]{\includegraphics[width=\x\textwidth, trim=120 0 140 10, clip=true]{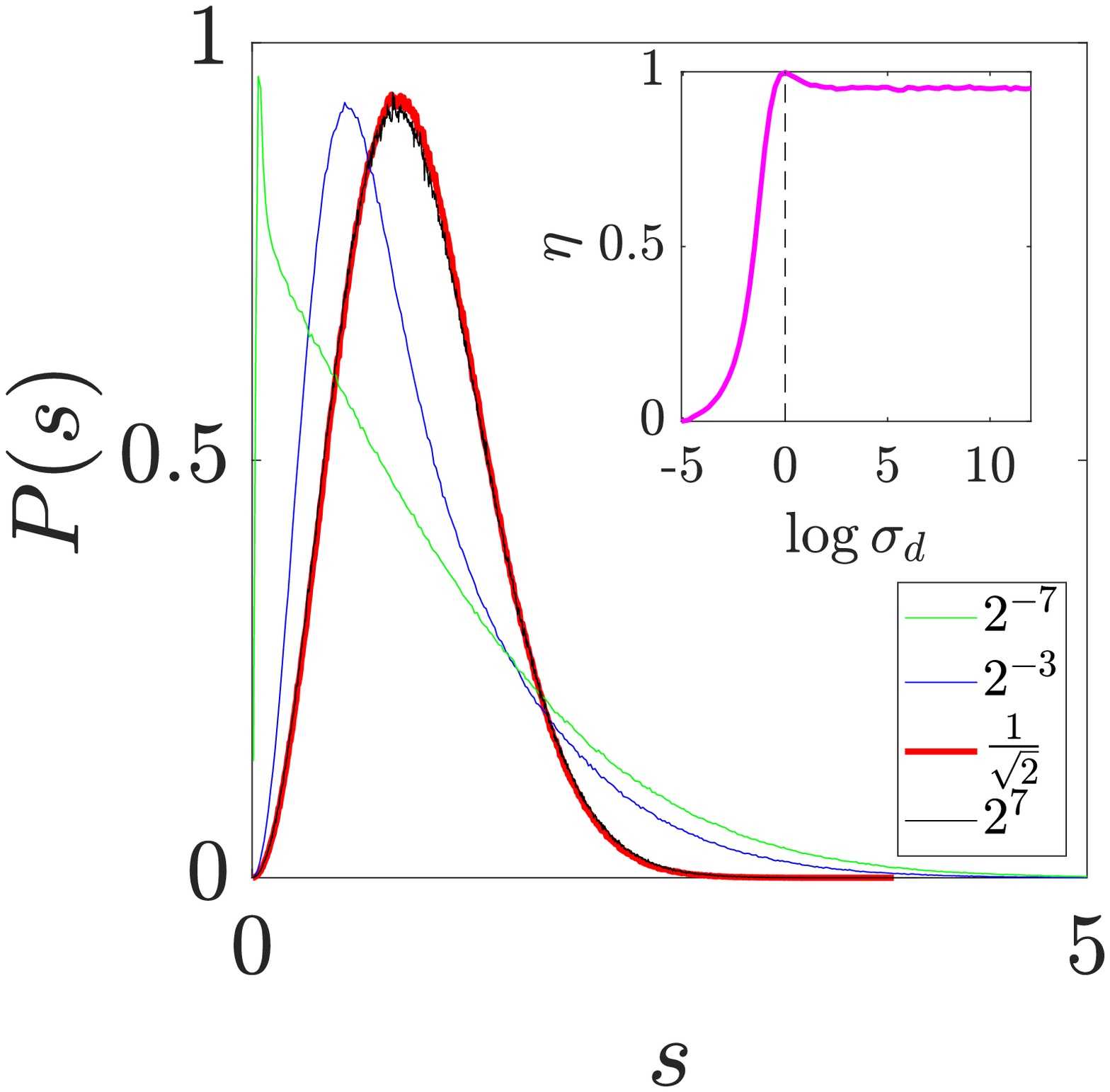}\label{ls_her_4_all}}
	\subfloat[][]{\includegraphics[width=\x\textwidth, trim=100 0 140 10, clip=true]{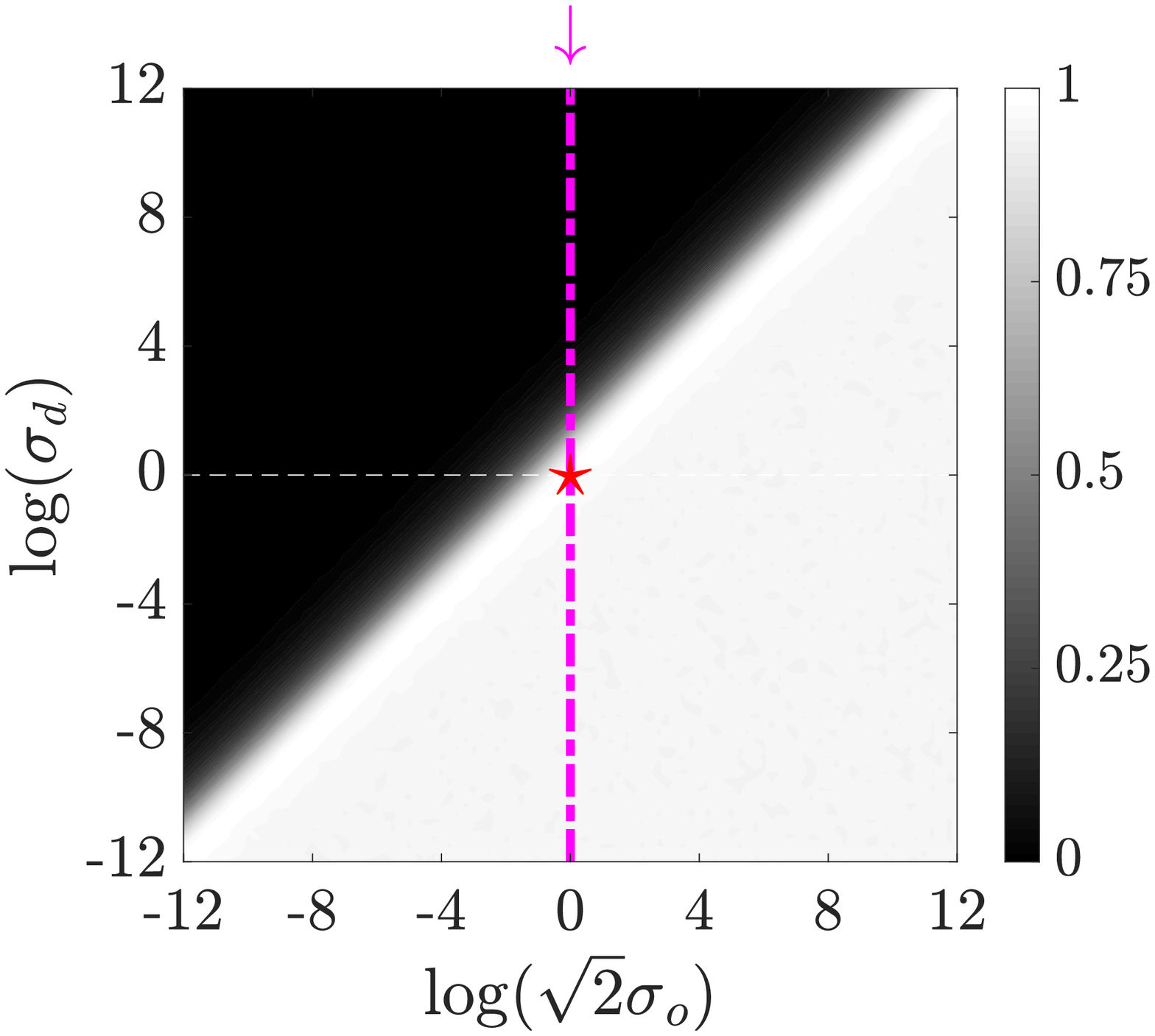}\label{pdiag_her_ls_4}}
	\caption[]{{\bf Transition from level clustering to repulsion in NNS of Hermitian matrices}: {\acrshort{pdf} of \acrshort{nns} for different $\gls{sdo}$
			with $\gls{sdd}=1$ for \subref{ls_her_3_all}~{$N=3$} \subref{ls_her_4_all}~{$N=4$}. Markers in \subref{ls_her_3_all} denote enumeration of Eq.~\eqref{eq_pdfls3her}.} 
		Insets in \subref{ls_her_3_all} and \subref{ls_her_4_all} show fitted transition parameter $\eta$ vs $\log(\gls{sdd})$ when $\gls{sdo}=1/\sqrt{2}$. Corresponding regions are denoted by bold lines in the \acrshort{pd}. The star in PD indicate GUE.
		{\acrshort{pd} constructed w.r.t. $\eta$ values} \subref{pdiag_her_ls_3} $N=3$, \subref{pdiag_her_ls_4} $N=4$.}
	\label{her_ls_3_4}
\end{figure}

\subsection{NNS distributions for $N = 3$}
Observing the form of $P({\bf E})$ in Eq.~\eqref{eq_herpdfe1e2} here also we suggest an  ansatz
	\begin{align}
	\label{eq_herpdfen}
	\begin{split}
	& \prob{\mathbf{E}} = C_0\exp\del{\sum_{i<j}^{N}\frac{1}{C}\del{\frac{1}{\gls{sdd}^2}-\frac{1}{2\gls{sdo}^2}}\del{\gls{Ei}-\gls{Ej}}^2}\exp\del{-\sum_{i=1}^{N}\frac{\gls{Ei}^2}{2\sigma_d^2}}\prod_{i<j}^{N}\del{|\gls{Ei}-\gls{Ej}|g\del{\frac{|\gls{Ei}-\gls{Ej}|}{\sqrt{C}}}}
	\end{split}
	\end{align}
$g(x)$ is defined in Eq.~\eqref{eq_herpdfe1e2}. For $\gls{sdd}=1,\gls{sdo}=1/\sqrt{2}$, Eq.~\eqref{eq_herpdfen} reduces to Eq.~\eqref{eq_pdfeng} 
the known expression for PDF of eigenvalues of GUE ({\it i.e.} $\beta=2$) \cite{Mehta1}.
We can evaluate the \acrshort{pdf} of \acrshort{nns} from
\begin{align}
	\label{eq_pdfls3her}
	\begin{split}
	&\prob{s} = C_0sg\del{\frac{s}{\sqrt{C}}}\int_{0}^{\infty}dy\Bigg[g\del{\frac{y}{\sqrt{C}}}g\del{\frac{s+y}{\sqrt{C}}}\exp\del{(2c_1-\frac{1}{3\gls{sdd}^2})(y^2+sy+s^2)}y(s+y)\Bigg],\quad \sbr{c_1=\frac{1}{C}\del{\frac{1}{\sigma_d^2}-\frac{1}{2\sigma_o^2}}}.
	\end{split}
\end{align}
Enumeration of this expression gives excellent match with corresponding simulated data with $C\approx{10.5}$ for $\gls{sdd}<\sqrt{2}\gls{sdo}$ and $C\approx{4}$ for $\gls{sdd}>\sqrt{2}\gls{sdo}$ Fig.~\ref{her_ls_3_4}\subref{ls_her_3_all}. 

In the limit $\gls{sdd}\gg\gls{sdo}$ we assume that the diagonal elements survive similar to the case of symmetric matrices and the PDF of NNS will also be given by $P_{cluster}(s)$ given in Eq.(\ref{eq_pdfpoissons}).
For the GOE limit, $\sigma_d=\sqrt{2}\sigma_o$, the PDF of NNS is given by
\begin{align}
	\label{eq_pdfs3gue}
	\probgue{s} &= \frac{3^{12}}{2^{28}\pi^4}s^2e^{-\frac{243}{64\pi}s^2}\del{144\sqrt{3}s(128\pi-81s^2)+e^{\frac{243}{256\pi}s^2}\del{3^9s^4-2^83^4\pi s^2+2^{14}\pi^2}Erfc\del{\frac{9\sqrt{3}}{16\sqrt{\pi}}s}}.
\end{align}
In the case {\gls{sdd}$\ll$\gls{sdo}} exact analytical expressions are hard to obtain as \gls{Hii} term survives in $\prob{\gls{H}}$ and we have to rely on numerical data. 
Here we observe that the \acrshort{pdf}s of NNS has close resemblance to that of \acrshort{gue}.
Thus, similar to the analysis for symmetric matrices, for intermediate statistics, we can construct a Brody like distribution. Limiting case of level clustering, $\prob{s}\sim\exp\del{-3s^2}Erfc(s)$ and level repulsion, $\prob{s}\sim\frac{2}{3\sqrt{\pi}}s^3(3-2s^2)\exp\del{-4s^2}+s^2\del{\frac{4}{3}s^2(s^2-1)+1}\exp\del{-3s^2}Erfc\del{s}$ suggest a transition function of the form $\prob{s}\sim\frac{2\eta}{3\sqrt{\pi}}s^3(3-2s^2)\exp\del{-4s^2}+s^{2\eta}\del{\frac{4\eta}{3}s^2(s^2-1)+1}\exp\del{-3s^2}Erfc\del{s}$. Thus for $\eta=1$ we will get clustering and $\eta=2$ produces quadratic level repulsion.
We need to normalize and unfold this function and the resultant expression is given in Appendix Eq.\ref{eq_her3phase1}.
Fig.~\ref{her_ls_3_4}\subref{pdiag_her_ls_3} depicts \acrshort{pd} for $3{\times}3$ Hermitian matrices obtained from estimation of $\eta$ by numerical fitting.
For $N=4$ numerical simulations indicate NNS distributions are similar to those of $N=3$ Hermitian matrices Fig.~\ref{her_ls_3_4}\subref{pdiag_her_ls_4}.

\captionsetup[subfigure]{position=top, labelfont=bf,textfont=normalfont,singlelinecheck=off,justification=raggedright}
\renewcommand{\x}{0.35}
\begin{figure}[!t]
	\vspace{-10pt}
	\centering
	\subfloat[][]{\includegraphics[width=\x\textwidth, trim=110 0 140 10, clip=true]{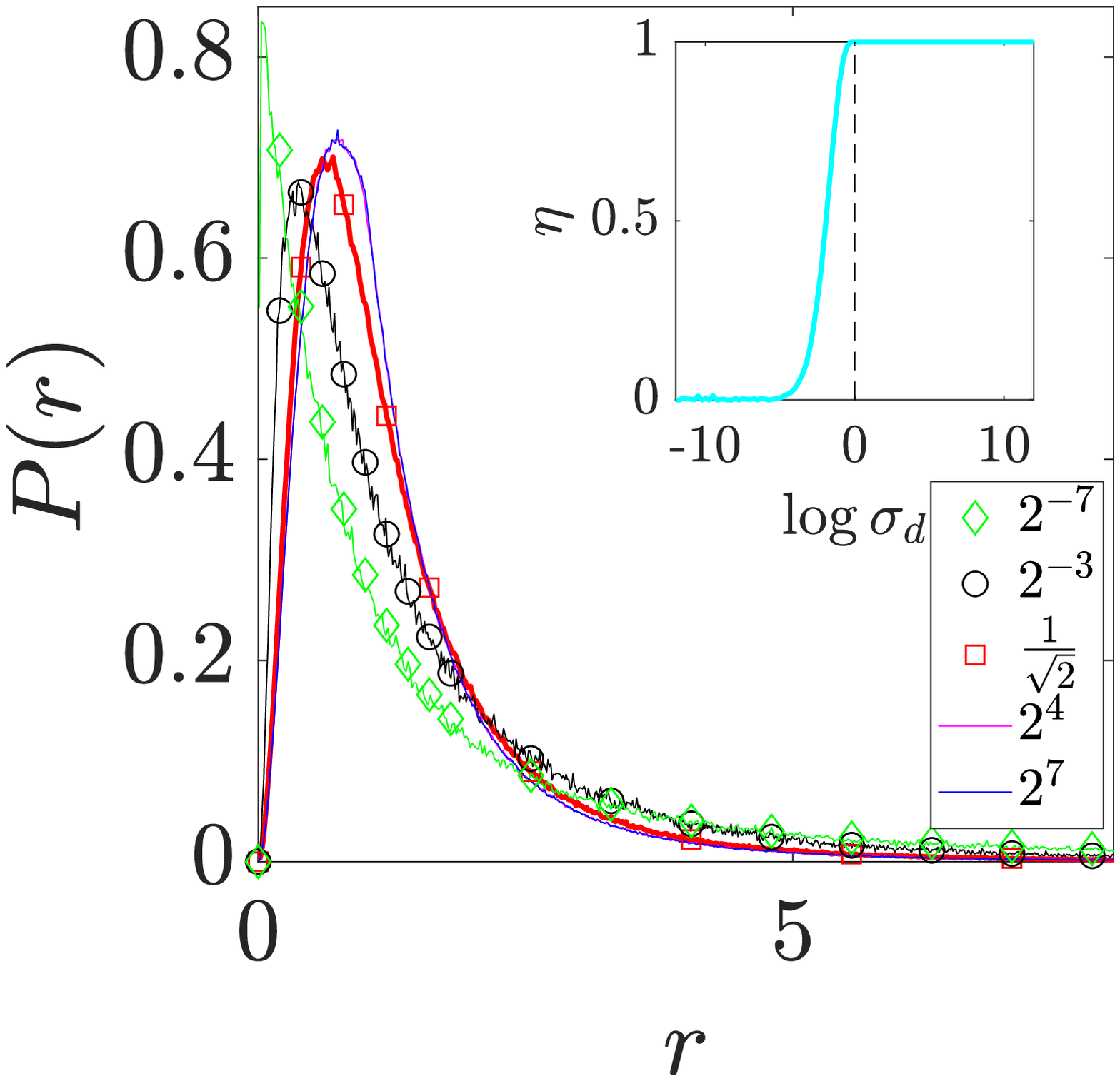}\label{rl_her_3_all}}
	\subfloat[][]{\includegraphics[width=\x\textwidth, trim=100 0 140 10, clip=true]{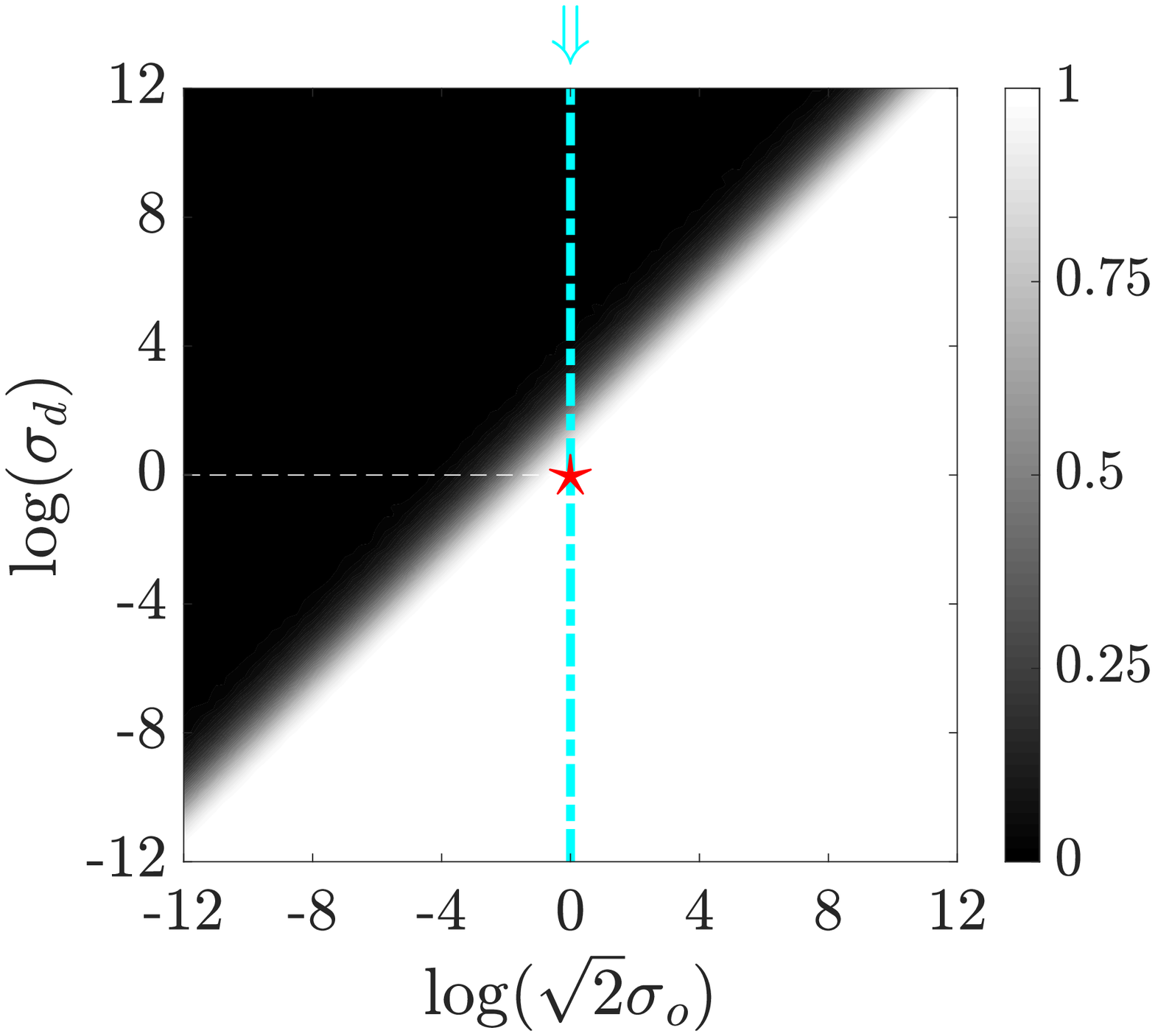}\label{pdiag_her_rl_3}}
	\vfill
	\vspace{-10pt}
	\subfloat[][]{\includegraphics[width=\x\textwidth, trim=110 0 140 10, clip=true]{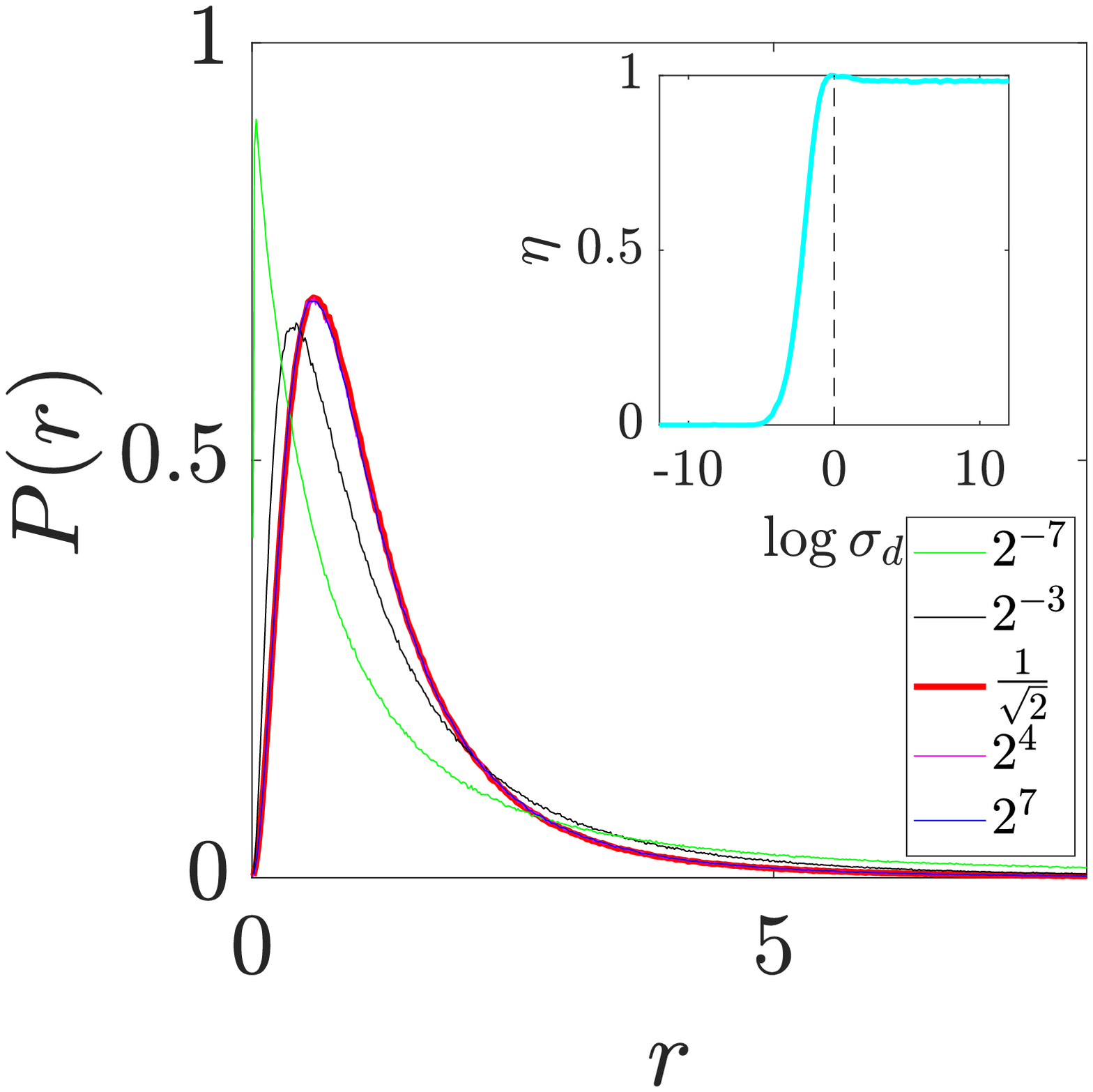}\label{rl_her_4_all}}
	\subfloat[][]{\includegraphics[width=\x\textwidth, trim=100 0 140 10, clip=true]{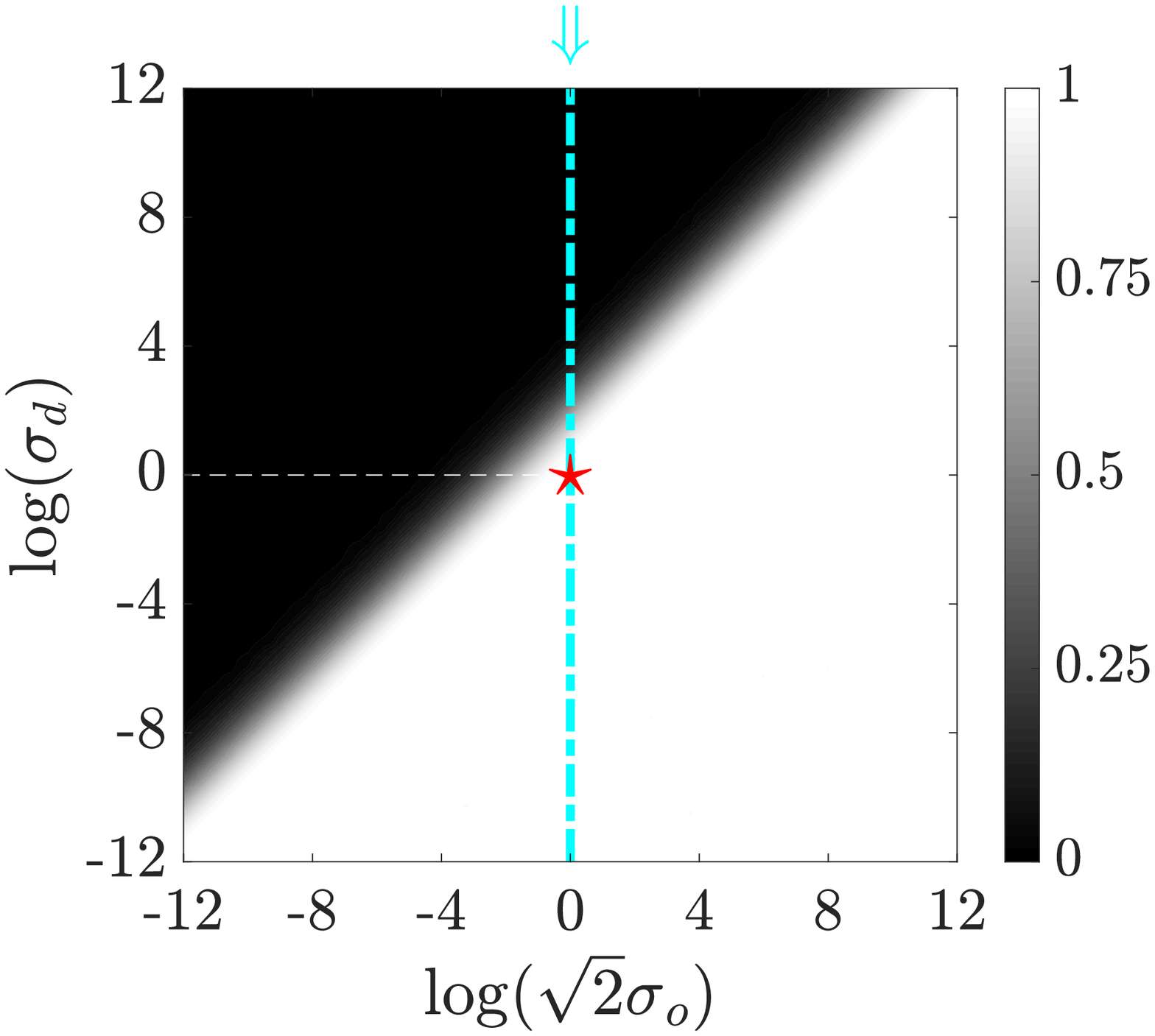}\label{pdiag_her_rl_4}}
	\caption[]{{\bf Transition from level clustering to repulsion in RNNS of Hermitian matrices}:
		{\acrshort{pdf} of \acrshort{rnns} for different $\gls{sdo}$
			with $\gls{sdd}=1$ for \subref{rl_her_3_all}~{$N=3$} \subref{rl_her_4_all}~{$N=4$}. Markers in \subref{rl_her_3_all} denote enumeration of Eq.~\eqref{eq_pdfrl3her}.} 
		Insets in \subref{rl_her_3_all} and \subref{rl_her_4_all} show fitted transition parameter $\eta$ vs $\log(\gls{sdd})$ when $\gls{sdo}=1/\sqrt{2}$. Corresponding regions are denoted by bold lines in the \acrshort{pd}. The star in PD indicate GUE.
		{\acrshort{pd} constructed w.r.t. $\eta$ values} \subref{pdiag_her_rl_3} $N=3$, \subref{pdiag_her_rl_4} $N=4$.}
	\label{her_rl_3_4}
\end{figure}

\subsection{RNNS distribution for $N=3$}
From the ansatz in Eq.~\eqref{eq_herpdfen} we obtain the RNNS distribution as follows
\begin{align}
	\label{eq_pdfrl3her}
	\begin{split}	
	&\prob{r} = C_0(r+r^2)\int_{0}^{\infty}dx\Bigg[g\del{\frac{(1+r)x}{\sqrt{C}}}g\del{\frac{rx}{\sqrt{C}}}g\del{\frac{x}{\sqrt{C}}}\exp\del{\del{2c_1-\frac{1}{3\gls{sdd}^2}}(1+r+r^2)x^2}x^4\Bigg].
	\end{split}
\end{align}
where $c_1$ is given in Eq.~\eqref{eq_pdfls3her}.  
For N=3, the RNNS distribution for GUE is known \cite{Atas2} and is given by
\begin{align}
	\probgue{r} &= \frac{81\sqrt{3}}{4\pi}\frac{\del{r+r^2}^2}{(1+r+r^2)^4}.
\end{align}
In the limit $\sigma_d\gg\sigma_o$, a Hermitian matrix is similar to a symmetric matrix and the PDF of RNNS is still given by Eq.~\eqref{eq_pdfpsn3r}. 
In the other limit, $\sigma_d\ll\sigma_o$, we could not find any analytical expression, however, numerical data suggests resemblance to GUE.
For $N=4$, an exact expression for PDF of RNNS for GUE has been exactly obtained \cite{Atas2}. In the limit $\sigma_d\gg\sigma_o$, PDF of RNNS is given by Eq.~\eqref{eq_pdfpsn4r} while for $\sigma_d\ll\sigma_o$ the numerical data also shows repulsion as in GUE. Based on these observations we use an additive interpolating function and estimate the transition parameter to obtain the phase diagrams shown in Fig.~\ref{her_rl_3_4}.

\section[Conclusion]{Conclusion}{\label{sect_con}}

We studied symmetric and Hermitian matrices where the diagonal and the off-diagonal elements are drawn from normal distributions but with different variances. 
This enables us to define a parameter $\tilde{\sigma} = \gls{sdd}/(\sqrt{2} \gls{sdo})$ which can be tuned to show that the spacing distributions of the eigenvalues show a crossover from level clustering to level repulsion.
In such generalized setting the symmetric matrices are not invariant with respect to orthogonal transformations and do not belong to GOE which is realized only for $\gls{sdd}=\sqrt{2} \gls{sdo}$.
The analytical calculation of $P(s)$ has been reported for $2 \times 2$ symmetric matrices \cite{Huu-Tai1} and 
we also observe that NNS statistics changes from clustering to GOE like repulsion.
We have proposed an ansatz for the eigenvalue distributions for $N \times N$ matrices and for $N=3$ we have derived an integral form of PDF of NNS as well as RNNS.
Here also we observe crossovers and obtain phase diagrams by invoking functions with parameters that distinctly identify clustering and repulsion. 
However, the phase diagram obtained for $N=3$ is qualitatively different from that of $N=2$. 
For $N=2$, the NNS distribution is symmetric for small and large values of $\tilde{\sigma}$ and the symmetry is reflected in the corresponding phase diagram.
In $\gls{sdd}-\gls{sdo}$ plane, GOE is in the center of a narrow band which shows repulsion while away from the band in either direction clustering exists.
On the other hand for $N=3$ the symmetry is broken and repulsion is observed only when $\gls{sdd} < \sqrt{2} \gls{sdo}$ which is also evident from PDFs of both NNS and RNNS.

In this work we have exactly solved for eigenvalue and spacing distributions of $2 \times 2$ Hermitian matrices similarly generalized and find the existence of crossover from clustering to repulsion.
Interestingly, $N=2$ generalized Hermitian matrices do not have the symmetry in its NNS distribution but the phase diagram shows clustering and repulsion as in GUE for $\gls{sdd} \sim \sqrt{2} \gls{sdo}$
but GOE like repulsion for $\gls{sdd} <<  \gls{sdo}$.
Introduction of relative variances thus results in spectral statistics to explore three distinct regimes.
For $N=3$ we also propose an ansatz and obtain an integral form of PDFs for NNS and RNNS and observe crossovers  
only to GUE like repulsion.

 \captionsetup[subfigure]{position=top, labelfont=bf,textfont=normalfont,singlelinecheck=off,justification=raggedright}
 \renewcommand{\x}{0.35}
 \begin{figure}[!t]
 	\vspace{-10pt}
 	\centering
 	\subfloat[][]{\includegraphics[width=\x\textwidth, trim=100 0 140 10, clip=true]{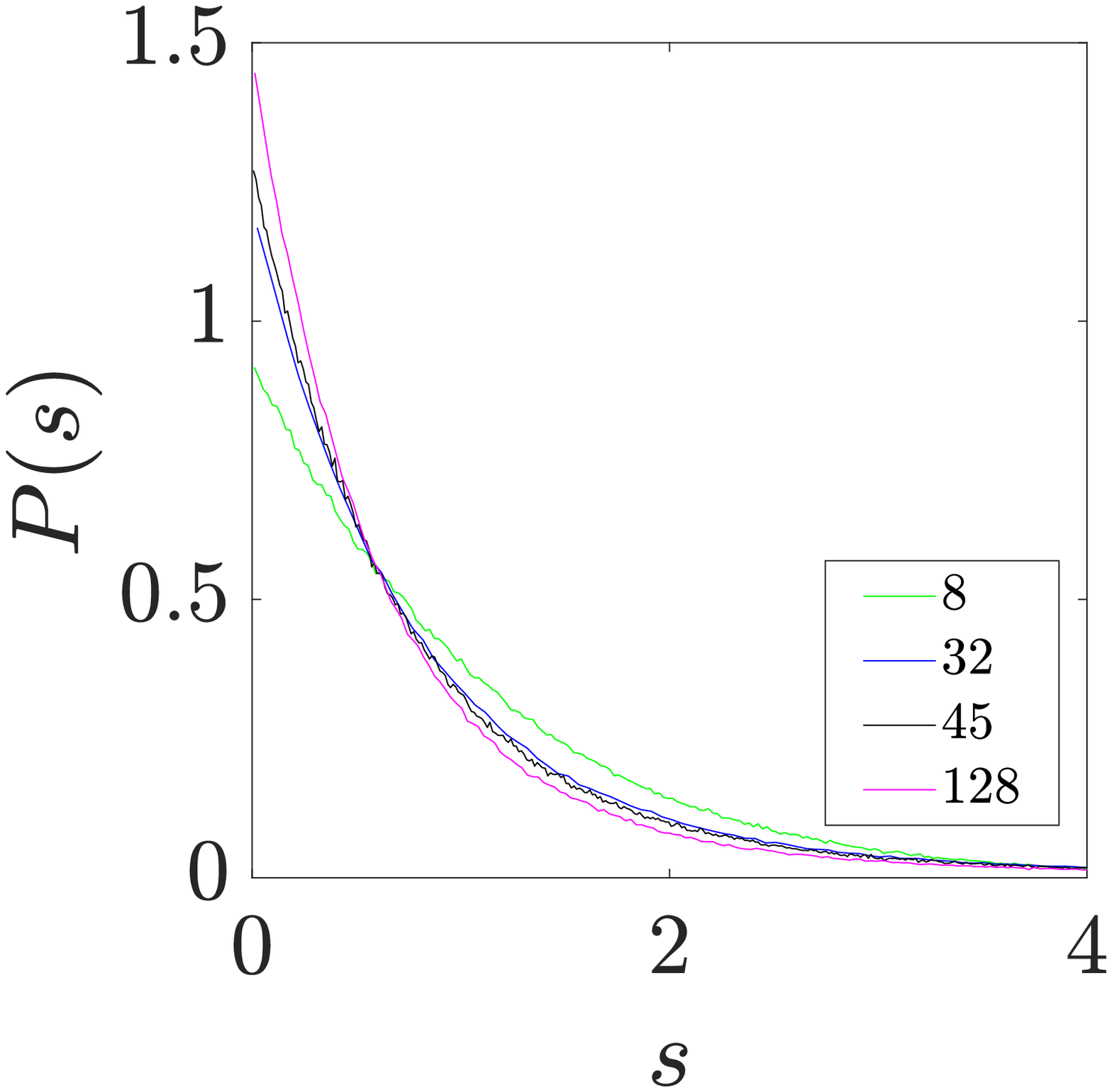}\label{sd_gg_so_sym}}
 	\subfloat[][]{\includegraphics[width=\x\textwidth, trim=110 0 140 10, clip=true]{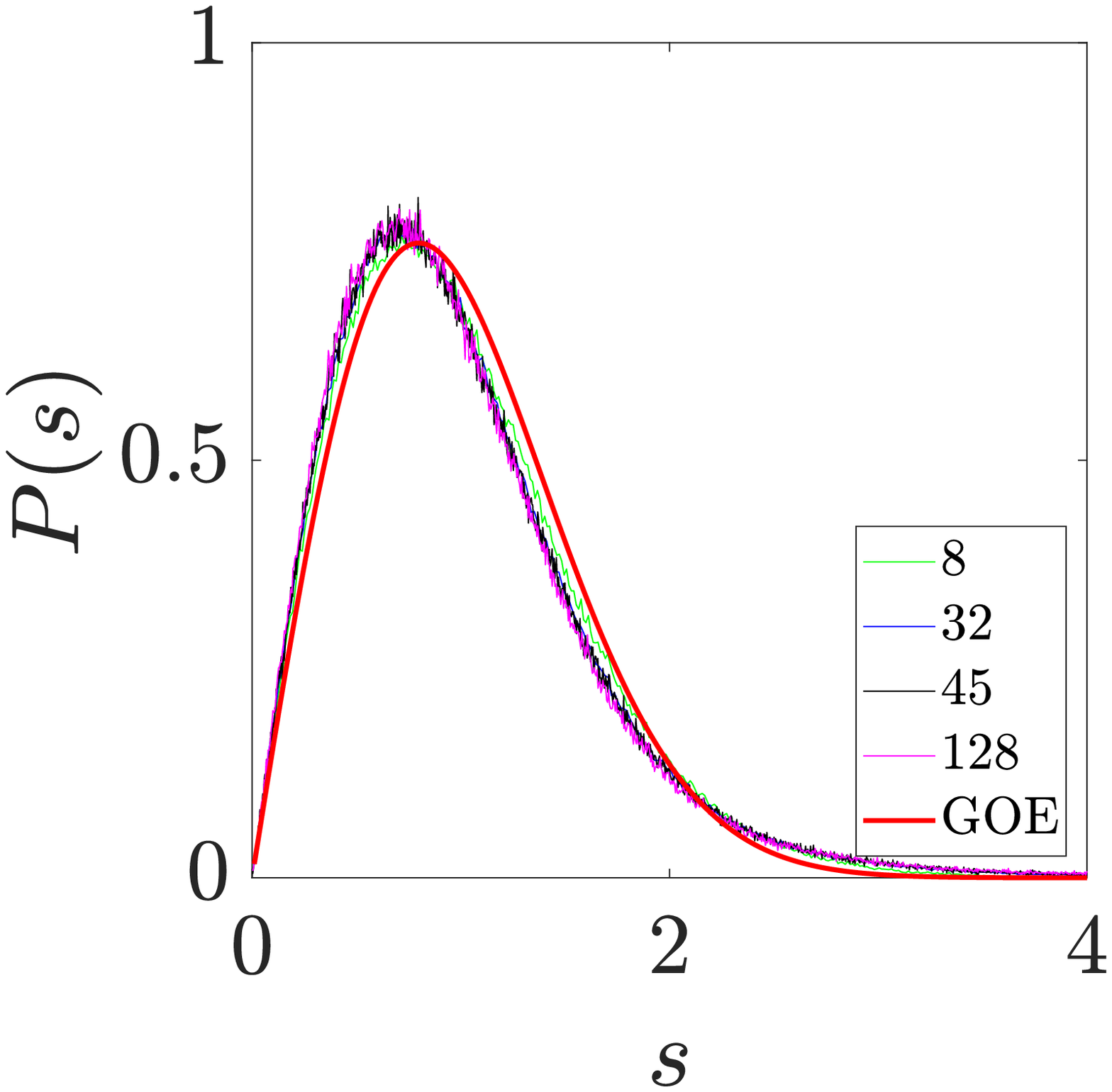}\label{sd_ll_so_sym}}
 	\vfill
 	\vspace{-10pt}
 	\subfloat[][]{\includegraphics[width=\x\textwidth, trim=100 0 140 10, clip=true]{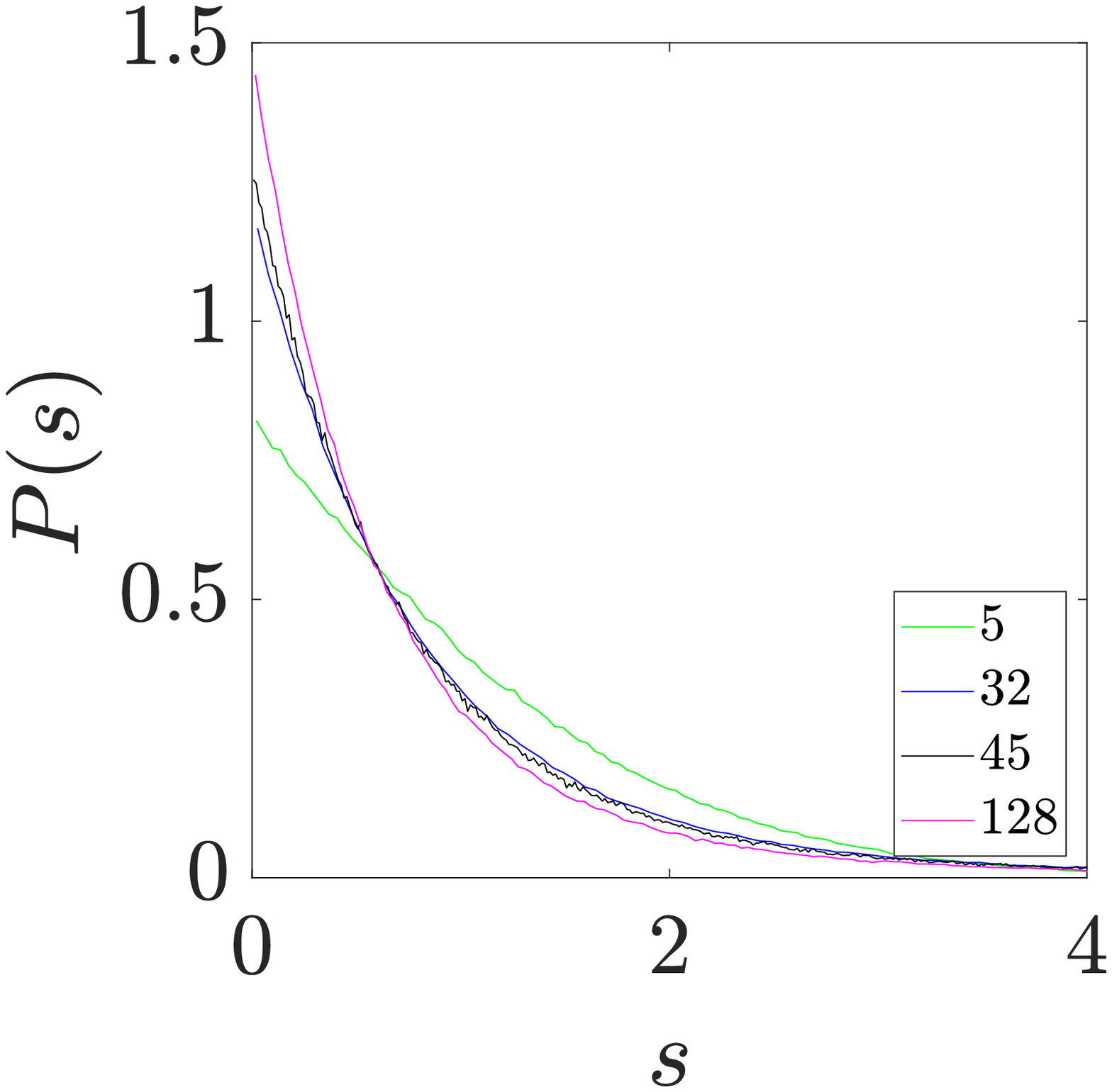}\label{sd_gg_so_her}}
 	\subfloat[][]{\includegraphics[width=\x\textwidth, trim=110 0 140 10, clip=true]{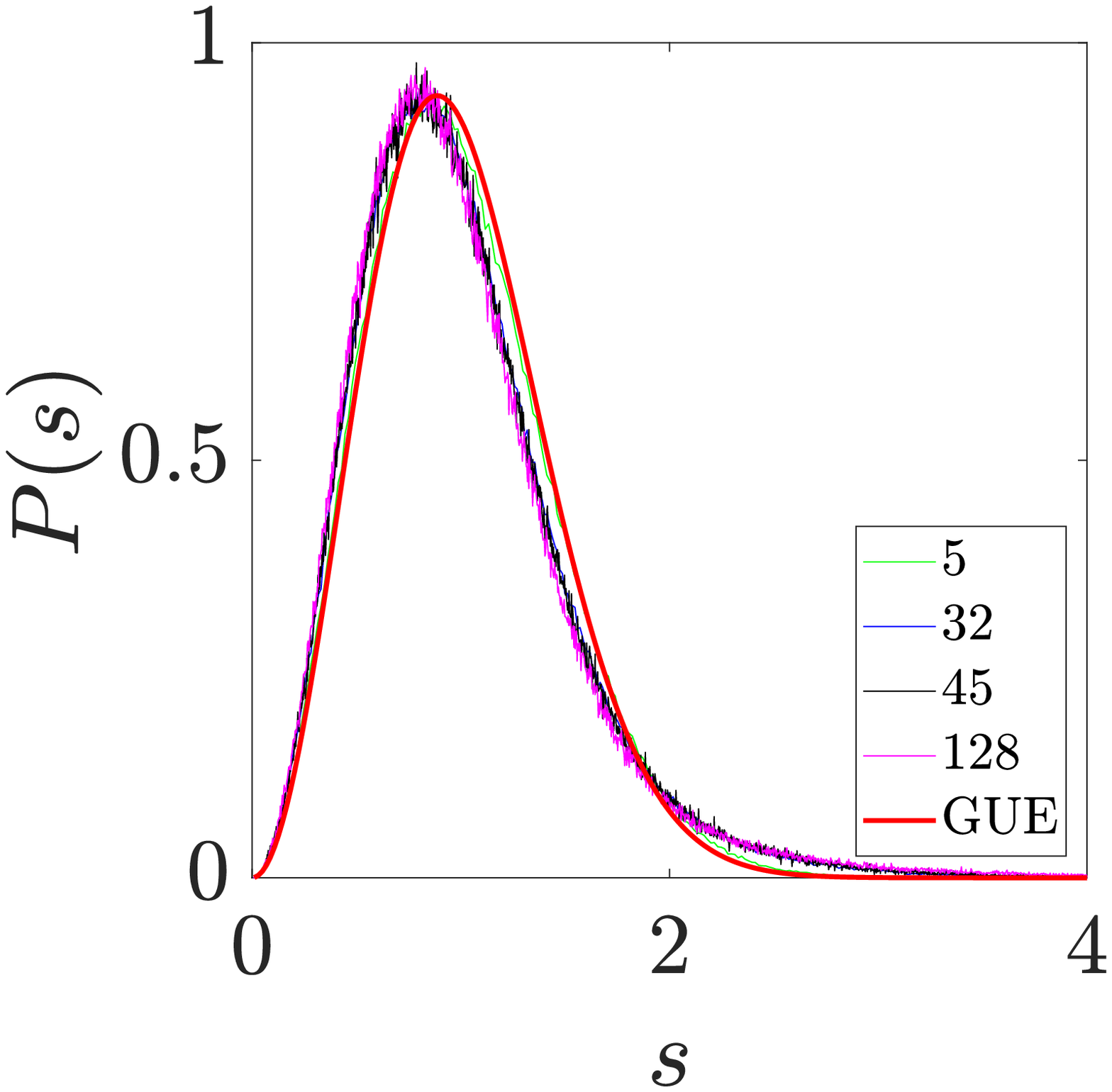}\label{sd_ll_so_her}}
 	\caption[]{{\bf PDF of NNS for different matrix sizes $N$}
 	\subref{sd_gg_so_sym} symmetric $\gls{sdd} >>  \gls{sdo}$
 	\subref{sd_ll_so_sym} symmetric $\gls{sdd} <<  \gls{sdo}$
 	\subref{sd_gg_so_her} Hermitian $\gls{sdd} >>  \gls{sdo}$
 	\subref{sd_ll_so_her} Hermitian $\gls{sdd} <<  \gls{sdo}$
 	.}
 	\label{sd_all}
 \end{figure}

The qualitative pictures evident from our calculations for $N=2,3$ matrices are also observed in numerical analysis of $N \times N$ symmetric and Hermitian matrices [Fig.~\ref{sd_all}].
For large $N$ the NNS distributions has resemblance to GOE (GUE) like repulsion for symmetric (Hermitian) matrices in the regime $\gls{sdd} \leq \sqrt{2} \gls{sdo}$ and clustering elsewhere.
The competition between relative strength of fluctuations in the on-site terms and the coupling terms can be possibly modelled as generalized random matrices resulting in intermediate statistics as observed in various theoretical studies and in experiments \cite{Rosenzweig1, Alt1, Abdul-Magd1, Robnik1}.
From a theoretical perspective it will be interesting to explore similar generalizations of matrices belonging to different symmetry classes and investigate the emergence of intermediate states and possibility of transitions.

\appendix
\section{}

\textbf{Normalized and Unfolded function used for interpolating $P(s)$ of Symmetric matrices, $N=3$:}
\begin{align}
	\label{eq_sym3phase1}
	\begin{split}
	&P\del{\eta; s} = \frac{\mu_\eta}{C_\eta}f(\mu_\eta{s}),\quad f(s)=\frac{2\eta}{\sqrt{\pi}}s^2\exp\del{-4s^2}-s^\eta(2\eta{}s^{2}-1)\exp\del{-3s^2}Erfc(s)\\
	&C_\eta = \frac{1}{2^{4+\eta}}\del{8\Gamma\del{1+\eta}{}_2FR_1\del{\frac{1+\eta}{2},\frac{2+\eta}{2},\frac{3+\eta}{2},-3}+\eta\del{2^\eta-4\Gamma\del{3+\eta}{}_2FR_1\del{\frac{3+\eta}{2},\frac{4+\eta}{2},\frac{5+\eta}{2},-3}}}\\
	&\mu_\eta = \frac{2^\eta\del{\eta+16\Gamma\del{\frac{3+\eta}{2}}\del{\frac{{}_2FR_1\del{\frac{2+\eta}{2},\frac{3+\eta}{2},\frac{4+\eta}{2},-3}}{2+\eta}-\frac{\eta\del{3+\eta}{}_2FR_1\del{\frac{4+\eta}{2},\frac{5+\eta}{2},\frac{6+\eta}{2},-3}}{4+\eta}}}}{\sqrt{\pi}\del{8\Gamma\del{1+\eta}{}_2FR_1\del{\frac{1+\eta}{2},\frac{2+\eta}{2},\frac{3+\eta}{2},-3}+\eta\del{2^\eta-4\Gamma\del{3+\eta}{}_2FR_1\del{\frac{3+\eta}{2},\frac{4+\eta}{2},\frac{5+\eta}{2},-3}}}}
	\end{split}
\end{align}
Where ${}_2FR_1$ is regularized Hypergeometric function.\\

\textbf{Normalized and Unfolded function used for interpolating $P(s)$ of Hermitian matrices, $N=3$:}
\begin{align}
	\label{eq_her3phase1}
	\begin{split}
	&\prob{\eta;s} = \frac{\mu_\eta}{C_\eta}f(\mu_\eta{s}),\qquad f(s)=\frac{2\eta}{3\sqrt{\pi}}s^3(3-2s^2)\exp\del{-4s^2}+s^{2\eta}\del{\frac{4\eta}{3}s^2(s^2-1)+1}\exp\del{-3s^2}Erfc\del{s}\\
	&C_\eta = \frac{1}{3\sqrt{\pi}2^{3+2\eta}}\eta\Bigg(4^\eta+24\sqrt{\pi}\fgamma{2\eta}{}_2F_1\del{\frac{1}{2}+\eta,1+\eta,\frac{3}{2}+\eta,-3}\\&\hspace{50pt}-4\sqrt{\pi}\fgamma{3+2\eta}{}_2F_1\del{\frac{3}{2}+\eta,2+\eta,\frac{5}{2}+\eta,-3}+\sqrt{\pi}\fgamma{5+2\eta}{}_2F_1\del{\frac{5}{2}+\eta,3+\eta,\frac{7}{2}+\eta,-3}\Bigg)\\
	&\mu_\eta = \frac{\sqrt{\pi}\del{189\eta{}4^\eta+128(27-4\eta+4\eta^3)\fgamma{2+2\eta}{}_2F_1\del{1+\eta,\frac{3}{2}+\eta,2+\eta,-3}}-16\eta(1+14\eta)\fgamma{\frac{3}{2}+\eta}}{1728\sqrt{\pi}2^{3+2\eta}C_\eta}
	\end{split}
\end{align}

\textbf{Symmetric matrix ($N=4$): in clustering limit $(\sigma_d\gg\sigma_o)$} normalized and unfolded PDF is given by,
\begin{align}
\label{eq_pdfpsn4s}
\probcluster{s} &= \frac{\mu}{C}\exp(-\frac{\mu^2}{4}s^2)\int_{0}^{\infty}dz\exp(-\frac{z^2}{4})Erfc\del{\frac{\mu{s}+z}{2\sqrt{2}}},\quad \mu\approx 0.732364,\quad C\approx 1.047198
\end{align}
Normalization constant, $C$ and mean, $\mu$ are numerically determined using the result \cite{Fayed1}
\begin{align}
\label{eq_erfnormal}
I(a,b,\infty) &= \frac{\sqrt{\pi}}{2}\int_{0}^{\infty}dte^{-t^2}Erf\del{at+b} = \frac{\pi}{4}Erf\del{\frac{b}{\sqrt{1+a^2}}}+\frac{\sqrt{\pi}}{2}e^{-b^2}\sum_{j=0}^{\infty}\frac{\del{\frac{a}{2}}^{2j+1}}{\fgamma{j+\frac{3}{2}}}H_{2j}(b)
\end{align}
where $H_j(x)$ is the $j^{th}$ Hermite polynomial. Then, for $N=4$, PDF of NNS in clustering limit is given by Eq.~\eqref{eq_pdfpsn4s1}.\\

\bibliography{rmt1_v1}
\bibliographystyle{ieeetr}

\end{document}